\documentclass[aip,jcp,preprint]{revtex4-1}
\usepackage[version=3]{mhchem} 
\usepackage{amsmath,amsthm,amssymb,amsfonts}
\usepackage{float}
\usepackage{siunitx}
\usepackage{graphicx}
\usepackage{booktabs}
\usepackage[english]{babel}
\usepackage{textgreek}
\usepackage{hyperref}
\usepackage[utf8]{inputenc}
\usepackage[T1]{fontenc}
\usepackage{mathptmx}
\usepackage{etoolbox}
\usepackage{dcolumn}
\usepackage{bm}
\usepackage{xcolor}
\usepackage{soul}

\DeclareSIUnit\angstrom{\text {Å}}
\newcommand{\beginsupplement}{%
    \setcounter{table}{0}
    \renewcommand{\thetable}{S\arabic{table}}%
    \setcounter{figure}{0}
    \renewcommand{\thefigure}{S\arabic{figure}}%
    \setcounter{equation}{0}
    \renewcommand{\theequation}{S\arabic{equation}}%
    \setcounter{section}{0}
    \renewcommand{\thesection}{S\arabic{section}}%
    \setcounter{subsection}{0}
    \renewcommand{\thesubsection}{S\arabic{subsection}}%
    \newcounter{SItab}
    \renewcommand{\theSItab}{S\arabic{SItab}}%
    \newcounter{SIfig}
    \renewcommand{\theSIfig}{S\arabic{SIfig}}%
}

\begin{document}


\title{Deconstructing the Origins of Interfacial Catalysis: Why Electric Fields are Inseparable from Solvation
}

\author{Solana Di Pino}
\affiliation{International Centre for Theoretical Physics (ICTP), Strada Costiera 11, 34151 Trieste, Italy}

\author{Debarshi Banerjee}
\affiliation{International Centre for Theoretical Physics (ICTP), Strada Costiera 11, 34151 Trieste, Italy}
\affiliation{Scuola Internazionale Superiore di Studi Avanzati (SISSA), Via Bonomea 265, 34136 Trieste, Italy}

\author{Marta Monti}
\affiliation{International Centre for Theoretical Physics (ICTP), Strada Costiera 11, 34151 Trieste, Italy}

\author{Gonzalo Díaz Mirón}
\affiliation{International Centre for Theoretical Physics (ICTP), Strada Costiera 11, 34151 Trieste, Italy}

\author{Giuseppe Cassone}
\affiliation{Institute for Chemical-Physical Processes, National Italian Research Council (CNR-IPCF), V.le F. Stagno d'Alcontres 37, 98158 Messina (Italy)}

\author{Ali Hassanali}
\email{ahassana@ictp.it}
\affiliation{International Centre for Theoretical Physics (ICTP), Strada Costiera 11, 34151 Trieste, Italy}




\begin{abstract}
In the last decade, there has been a surge of experiments showing that certain chemical reactions undergo an enormous boost when taken from bulk aqueous conditions to microdroplet environments. The microscopic basis of this phenomenon remains elusive and continues to be widely debated. One of the key driving forces invoked are the specific properties of the air-water interface including the presence of large electric fields and distinct solvation at the surface. Here, using a combination of classical molecular dynamics simulations, the chemical physics of solvation, and unsupervised learning approaches, we place these assumptions under close scrutiny. Using phenol as a model system, we demonstrate that the electric field at the surface of water is not anomalous or unique compared to bulk water conditions. Furthermore, the electric field fluctuations de-correlate on a timescale of $\sim$10 ps implying that their role in \emph{activating} much slower chemical reactions remains inconclusive. We deploy a recently developed unsupervised learning approach, dubbed information balance, which detects in an agnostic fashion the relationship between the electric field and solvation collective variables. It turns out that the electric field on the hydroxyl group of the phenol is mostly determined by phenol hydration including the proximity and orientation of nearby water molecules. We caution that the growing attention of the role that electric fields have garnered in enhancing chemical reactivity at the air-water interface, may not reflect their actual importance.
\end{abstract}

\maketitle


\section{Introduction} \label{sec:intro}

The structural, dynamical and dielectric properties of aqueous interfaces have been implicated for a wide range of problems ranging from atmospheric chemistry\cite{limmer2024,jungwirth2006} to the origins of life\cite{vaida2021prebiotic,lopezfranciscoreview2020}. Over the last decade or so there have been a flurry of studies suggesting that chemical reactions can be significantly accelerated near water interfaces compared
to when they occur in the bulk\cite{lopezfranciscoreview2020}. Sharpless and co-workers coined the term `on-water' catalysis observing that some organic chemistry reactions
dramatically speeded up when conducted in water suspensions compared to organic solvents as well as in neat conditions\cite{sharpless2005}. The consistent experimental observation
is that moving from bulk water to microdroplet conditions, leads to the acceleration of certain chemical reactions\cite{zare2018,zareangewandte2015,cooks2011,Lee_Banerjee_Nam_Zare_2015,gordon2025,zheng2025}. This enhancement is thought to occur at the boundary
between water and air where the chemical reactants may experience different underlying physical driving forces. The molecular origins that lead to the
enhanced catalysis remains a topic of lively debate in the literature.

There have been several factors that are currently being examined in the community to rationalize the origins of the enhanced chemical reactivity including the presence of large electric fields and their fluctuations (ranging
between $10^{9}-10^{10}$ V/m)\cite{field12023,field22020,martinscosta2023,xiong2020,Hao2022,chan2025}, interfacial solvation and anisotropy of chemical reactants induced by the air-water boundary\cite{martins2023electrostatics,shibdas2023,cookssolvation2020}, curvature of the interface \cite{delapuente2024}, the surface activity of protons or hydroxide ions\cite{colussi2021,BANERJEE2023117024,chan2025} and finally,
geometric confinement effects\cite{Lee_Banerjee_Nam_Zare_2015,zhang2022}. A particular reaction that has caught significant attention is the one leading to the formation of hydrogen-peroxide ($H_2O_2$) in microdroplets
observed by the Zare group\cite{zare2019}. Over the past few years, the Mishra group has shown through a series of careful and systematic studies that the high yields of $H_2O_2$ originally observed by the Zare group arise not from the presence of intrinsic fields at the interface, but rather from impurities such as ozone or dissolved oxygen\cite{mishra2021,mishra2022,mishra2024,eatoo2025}. The importance of dissolved oxygen as a source of both $H_2O_2$ and hydroxyl radicals has also recently been pointed out by other studies\cite{asserghine2025}. On the other hand, the Head-Gordon group using molecular simulations, have implicated
large electric fields and partial solvation as playing a key role in the creation of hydroxyl radicals at the air-water interface\cite{gordonohradical2022}.

The potential role of interfacial electric fields is partly rooted on the evidence that \emph{strong external static} electric fields are sometimes capable of producing catalytic effects. In fact, experimental evidence has shown that electrostatic potential gradients can selectively enhance and catalyze Diels-Alder reactions~\cite{Aragones} and other industrially significant transformations~\cite{Huang}. Complementary insights into the effects of strong electric fields have come from quantum chemical computations on isolated molecules \cite{Shaik1,Shaik2,Shaik2016} and from extensive simulations on condensed-phase systems \cite{Cassone_ChemSci17,Cassone_TopCatal}. These strong external artificial fields on the order of $\sim10^{9}-10^{10}$ V/m align in magnitude with the local electric fields naturally occurring in condensed matter \cite{Saggu,Boxer,Fried_15,samantaray2025} and are therefore thought to introduce catalytic activity. A recent study by Xie and coworkers showed using \emph{ab initio} molecular dynamics simulations coupled with free energy calculations, that the barrier for a certain Diels-Alder reaction was actually slightly higher at the air-water interface and that the intrinsic fields and their fluctuations did not help speed up the reaction\cite{jing2024jacs}. In the same sense, Bonn's group has carried out surface sensitive experiments that question the strength of electric fields and the time scale of their fluctuations\cite{shirley2025}.

The rate of a chemical reaction is typically determined by the slowest degree of freedom that shapes the lowest free energy path
going from the reactant to the product\cite{laioparrinello2002}. For a chemical process that occurs in solution, solvation can play an important role in, e.g., stabilizing the transition-state
or facilitating charge-transfer events such as proton or electron transfer\cite{hassanali2011,bagchi1989,mundy2018}. It is known that the fluctuations of molecular dipoles in liquid water can produce fields larger than $10^{10}$~V/m~\cite{Geissler}, whereas in aqueous solutions~\cite{Chalmet_JCP2001,Saikally_PNAS2005,Ruiz-Lopez_PNAS2021,Cassone_JPCL23} and in presence of solvated ions~\cite{Sellner,Kathmann_check} local field intensities exceeding $\sim3\times10^{10}$~V/m have been reported. Besides, electric field strengths on the order of $10^{9}$ V/m have been suggested to persist at the air-water interface~\cite{xiong2020,Zare_PNAS2023,martins2023electrostatics}. If these electric fields play a central role in chemical reactions, the timescales associated with their fluctuations must also be taken into consideration. In addition, since electric fields are ultimately rooted in charge density distributions arising from the water molecules oriented at the interface, can they really be de-coupled from solvation?

In this work, we tackle the question of the interplay between electric fields and solvation using phenol as a test case. Electric fields at interfaces and how they modulate surface potentials has been a very active area of theoretical investigation using both classical and first principles approaches\cite{kathamnn2008,kathmann2011}. Our choice of using phenol is rooted in the fact that the phenolic chemical group is abundantly found in organic matter and plays a crucial role in the creation of aerosols\cite{hendrix2024}. Furthermore, due to its complex acid-base chemistry, it can also undergo various types of chemical reactions\cite{hendrix2024}. Tahara and co-workers recently demonstrated that the photodissociation of phenol occurs several orders of magnitude faster than in the bulk\cite{Kusaka2021}. The precise origins of this effect remain unknown, although it is very likely tuned by specific details of how the ground and excited-state potential energy surfaces evolve differently for phenol in the bulk and at the interface, as recently shown by Morita\cite{morita2022}. Our goal here is
to use this system as a simple model to investigate how electric fields on the hydroxyl moiety of the phenol change moving from bulk to interfacial environments and to examine the connection between solvation and electric fields. We want to accomplish this using an agnostic, unsupervised learning approach based on the Information Imbalance (II) introduced by Laio and collaborators \cite{Glielmo_PNASNexus_2022}. This method allows us to directly comment on which set of collective variables are more informative in a given system.

Using classical molecular dynamics simulations of phenol in different environmental conditions, we show that the electric fields along the hydroxyl group of the
phenol are no different in bulk or interfacial conditions. In fact, in our current models, the electric field on a water molecule in the bulk is actually slightly larger than that of the phenol at the interface. For both the situations where the phenol is in the bulk and at the interface, we observe large fluctuations in the electric field. However, the timescales associated with these fluctuations are rather fast, ranging between $\sim1-10$ ps. This suggests to us at least, that if electric field fluctuations are to play a role in speeding up chemistry, the reaction must occur on a similar timescale and its reaction axis must be properly aligned with the field direction within this time-frame. In fact, it is well-known that, for externally applied static electric fields, both the magnitude and the orientation determine the fate of a chemical reaction toward either catalysis or inhibition \cite{Shaik2016,Zhao_2025}. By examining the solvation structure of water around the phenol using the II test, we demonstrate that the electric field on the phenol is essentially described by the proximity and orientation of the nearby solvent molecules. These findings very closely mirror those of Ruiz-López and co-workers who, using QM/MM simulations, examined the coupling between the field on a hydroxyl ion and solvent reorganization\cite{martinscosta2023}.

The paper is organized as follows. We begin in Section \ref{sec:methods} by a review of the Methods employed in our work. Subsequently, we move to the Results starting in Section \ref{subsec:efield} with the analysis of the static properties of the electric field on the phenol in the different systems studied. In Section \ref{subsec:solvation} and Section \ref{subsec:ii} we examine the coupling between the electric field and solvation and finally we conclude in Section \ref{sec:discussion} with our perspectives.

\section{Methods} \label{sec:methods}

\subsection{Molecular Dynamics Simulations}

\begin{figure}[ht!]
  \centering
  \includegraphics[width=0.80\linewidth]{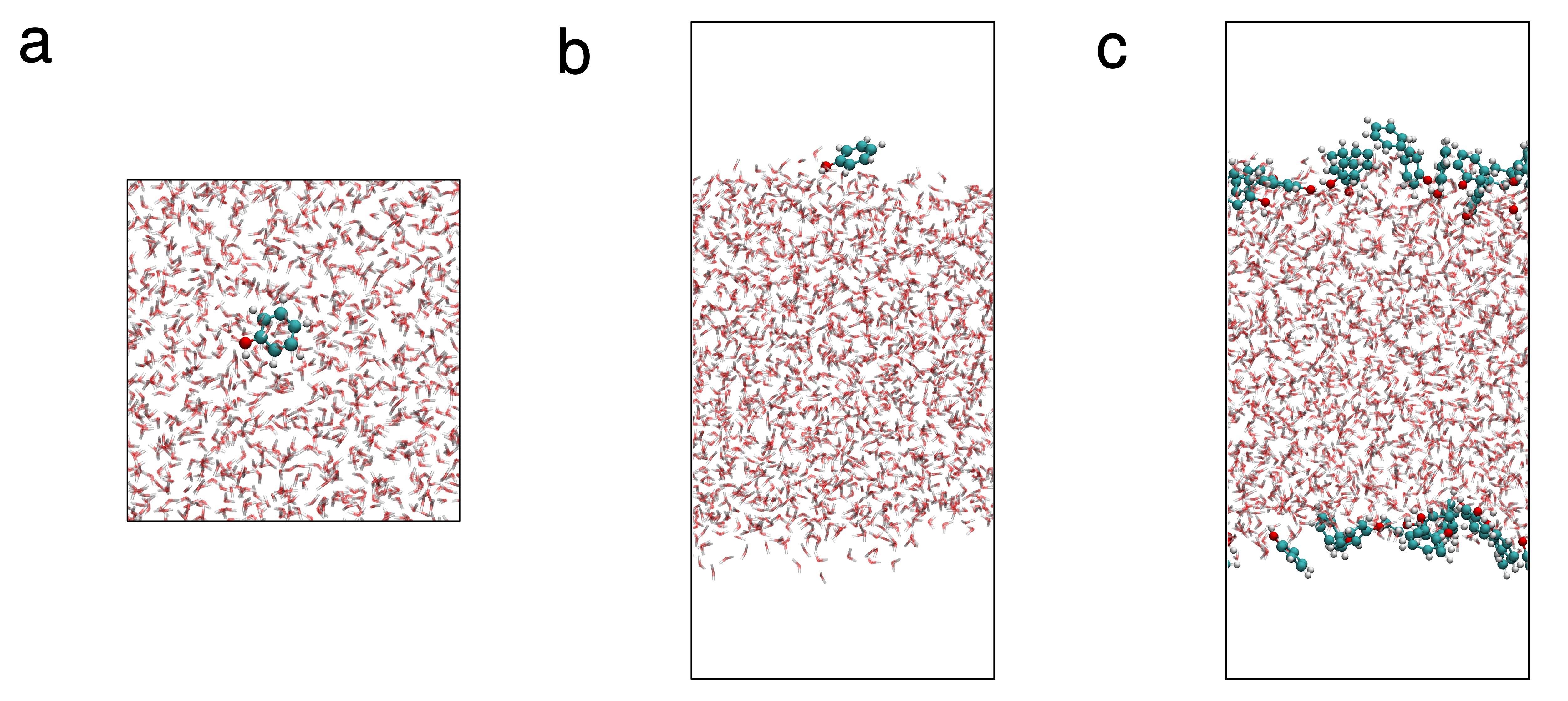}
  \caption{Snapshots of the three systems studied in this work: a) phenol in the bulk, b) one phenol molecule at the air/water interface (1PHX) and c) 25 phenol molecules at both interfaces of a water slab (25PHX).}
  \label{fig:systems}
\end{figure}

We carried out classical molecular dynamics simulations of three systems: bulk water with a single phenol molecule in it (bulk), a water slab with a phenol molecule at the air/water interface (1 PHX) and a water slab with 25 phenol molecules at each air/water interface of the slab, with a total of 50 phenol molecules in the whole system (25 PHX). We chose this concentration of phenol molecules which results in a surface excess of phenol of approximately 0.7 molecules per nm$^{2}$ corresponding to the lower concentration limit employed in the experimental work by Tahara on the photodissociation of phenol at the air/water interface\cite{Kusaka2021}. Snapshots of the 3 simulated systems are presented in Figure\ref{fig:systems}. This system also serves as a good limiting case for probing the effect of aggregation/clusters at the air-water interface on the electric fields.

The various systems were prepared as follows. The bulk system consisted of 66054 water molecules and one phenol molecule (bulk), with a simulation box of 128 {\AA} x 12 {\AA} x 12 {\AA} (obtained after NPT equilibration, see below). For the slab systems, we used a previously equilibrated water box containing 3636 water molecules with dimensions of 60 {\AA} x 60 {\AA} x 140 {\AA}, as reported in Ref.\cite{Donkor_JCTC_2023}. To this slab, either one (1PHX) or 25 (25PHX) phenol molecules were added to each side of the interface.

Following system preparation, simulations were carried out in multiple stages. For the bulk system, we first performed energy minimization, followed by a 50 ns equilibration in the NPT ensemble at 1 atm and 300 K to stabilize the density. A 100 ns production run was then performed in the NVT ensemble. For the slab systems, energy minimization was conducted after the phenol molecules were added, followed by a 100 ns production simulation in the NVT ensemble.

All simulations were conducted using a Langevin thermostat set to 300 K with a damping constant ($\gamma$) of 2.0 ps$^{-1}$ and a time step of 2 fs. Periodic boundary conditions were used in all directions for all systems, for the interface systems the z-axis (normal to the interface) was large enough to ensure no interactions between periodic boundaries in that direction. The OPC water model\cite{OPC, OPC3} was employed since it has been shown to accurately predict the surface tension of water. Specifically, at room temperature OPC yields a surface tension of 75 mN/m in excellent agreement with the experimentally measured 70 mN/m \cite{OPC_surface_tension}. Parameters for phenol were generated using the LEaP protocol within the AMBER package. Long-range electrostatic interactions were treated using the Particle Mesh Ewald (PME) method with a cutoff of 1 \AA. Simulations were performed using the AMBER\cite{ambertools2023} and GROMACS\cite{gmx_1995, gmx_2001, gmx_2005, gmx_2008, gmx_2013, gmx_2015_article, gmx_2015_conference} software package. The latter was used since it allows for more convenient postprocessing tools to extract the surface tension in order to validate the models of the interfacial systems which due to the low concentration (surface access) are basically the same as for the OPC neat air-water interface.

In this work, we calculated the electric field at the middle of the O-H bond of the phenol molecule. This was determined considering only the electric field produced by all the water molecules in the box for the system consisting of only one phenol. In the case of the system of 50 phenol molecules, the contribution of all the phenol molecules was also factored in addition to the water contribution, excluding the self contribution. The electric field was calculated according to Equation \ref{eq:ef}, as done in previous studies\cite{Reischl20022009,jong2018}, using the atomic point charges given by the OPC force field, used to describe the water molecules. Since the GAFF force fields used to describe the interaction in these systems don't include polarization and charge transfer effects we validated this methodology using the QM/MM approach that is described in the SI text.

\begin{equation}\label{eq:ef}
    \bar{E} = \frac{1}{4\pi\epsilon_0} \sum_{i=1}^N \frac{q_i}{|\pmb{r}_i|^2} \hat{\pmb{r}_i}
\end{equation}

In this equation $\pmb{r}_i$ is the distance of the charge to the midpoint of the O-H bond and N is the total number of water molecules in the system. The field obtained at the center of the O-H bond can be subsequently projected along the O-H bond direction yielding a three component vector, according to equation \ref{eq:proj}, used for the ensuing analysis that will be described in more detail later.

\begin{equation}\label{eq:proj}
    proj\pmb{E} = (\pmb{E} \cdot \hat{\pmb{r}}) \hat{\pmb{r}}
\end{equation}


\subsection{Information Imbalance (II)}

In recent years machine learning has become an integral part of both conducting and understanding molecular simulations\cite{jackson2023}. It is rather well appreciated that most molecular systems need to be described in terms of fluctuations occurring in high dimensional free energy landscapes\cite{laio2021}. There is thus a lot of effort being placed into developing methods that help quantify, probe, and interpret the complexity that is inherent in molecular systems. In this particular work, we explore the relationship between electric fields and the solvation of phenol. Specifically, we re-cast this problem within the language of a statistical test known as the information imbalance\cite{Glielmo_PNASNexus_2022} (II), which seeks to quantify the relationship between different distance measures applied to the same data. Within the context of our work here, we ask whether the electric field on the phenol can be rationalised solely, or at least mostly, by solvation coordinates. We begin by first briefly reviewing the II theory.

II is a measure to compare the information content between two distance measures $d_A$ and $d_B$, defined over a given data set. Within our context $d_B$ would be the distances in the electric field vector while $d_A$ will be the distances with respect to solvation degrees of freedom. Then, distances $d_B$ can be defined between points $i$ and $j$ (which are two different values of the electric field vector components at two times in our trajectory) as: $d_B^{ij} = ||E_x^i - E_x^j|| + ||E_y^i - E_y^j|| + ||E_z^i - E_z^j||$, where $||.||$ denotes the Euclidean norm. Distances in $d_A$ are defined similarly using a set of solvation coordinates that we will define below in the text.

Given these distances, one refers to $r^{ij}_A$ the distance rank of point $j$ with respect to point $i$ according to $d_A$. Similarly, $r^{ij}_B$ is the distance rank of $i$ with respect to $j$ according to distance $d_B$. In this way, one can say that $d_A$ is informative with respect to $d_B$ if points close according to $d_A$ are also close according to $d_B$. It is possible to say that $d_A$ is informative with respect to $d_B$ when points $ij$ that are close neighbors according to $d_A$ remain close also according to $d_B$.

The II from $d_A$ to $d_B$ is defined as:

\begin{equation}\label{eq:iib}
    \Delta(d_A \rightarrow d_B) := \frac{2}{N} \langle r_B \mid r_A\, \leq k \rangle = \frac{2}{N^2\,k} \sum_{i,j:\, r^{ij}_A\, \leq k} r^{ij}_B\,,
\end{equation}

where $N$ is the total number of points in the data set and $k$ is the number of nearest neighbor points that are considered. Eq.\ref{eq:iib} gives us the II, represented as $\Delta(d_A \rightarrow d_B)$, as a number between 0 and 1. $\Delta(d_A \rightarrow d_B) =0$ occurs when all points that are nearest neighbors according to $d_A$ remain nearest neighbors in $d_B$, whereas $\Delta(d_A \rightarrow d_B) =1$ occurs when the nearest neighbors according to $d_A$ are randomly distributed in $d_B$. The former case is when $d_A$ is maximally informative with respect to $d_B$, and the latter is when it is minimally informative.

Recently, an extension to the II has been proposed\cite{Wild_NatComm_2025}, where each feature that makes up a given distance measure is scaled by a weight, and then these weights are optimized through gradient descent. This formalism allows us to combine heterogeneous features, which might have very different units of measurement, in a straightforward manner, without needing to worry about scaling the data appropriately. This is referred to as the Differentiable Information Imbalance (DII) and it allows us to determine the most informative subset of coordinates or features to define a distance metric within a given data set. The DII is defined as:

\begin{equation}\label{eq:dii}
    DII(d_A(\pmb{w}) \rightarrow d_B) := \frac{2}{N^2} \sum_{i,j=1}^{N} c_{ij} (\lambda, d_A(\pmb{w}))r^{ij}_B
\end{equation}

where,
\begin{equation}
        c_{ij}(\lambda , d_A(\pmb{w})) := \frac{e^{-d^{ij}_A(\pmb{w})/\lambda}}{\sum_{m\neq i} e^{-d_A^{im}(\pmb w)/\lambda}}
\end{equation}

Here, $\pmb{w}$ are variational parameters to be optimized via gradient descent in order to assign different weights to different features in distance $d_A$, and the parameter $\lambda$ is chosen according to the average and minimum nearest neighbor distances. In this way, it is possible to both automatically identify the most relevant features in distance $d_A$ that are maximally predictive with respect to distance $d_B$, as well as the respective ratios in which they should be combined. In the limit $\lambda \rightarrow 0$, the DII (Eq. \ref{eq:dii}) is equivalent to and can be viewed as a continuous version of the standard Information Imbalance (Eq. \ref{eq:iib}).

In this work, we primarily applied the DII to determine the information content of a set of solvation-related features on the electric field vector along the O-H bond of the phenol molecule. To perform a feature search and to avoid the problem of combinatorial explosion that would naturally arise if one were to test every possible feature subset, we use the forward greedy search algorithm introduced in Ref.\citenum{Glielmo_PNASNexus_2022} coupled with the DII approach that is implemented in the Python package DADApy \cite{glielmo2022dadapy}.


\section{Results}
\label{sec:results}

\subsection{Phenol Electric Field: Statics and Dynamics}\label{subsec:efield}

\begin{figure}[ht!]
  \centering
  \includegraphics[width=0.95\linewidth]{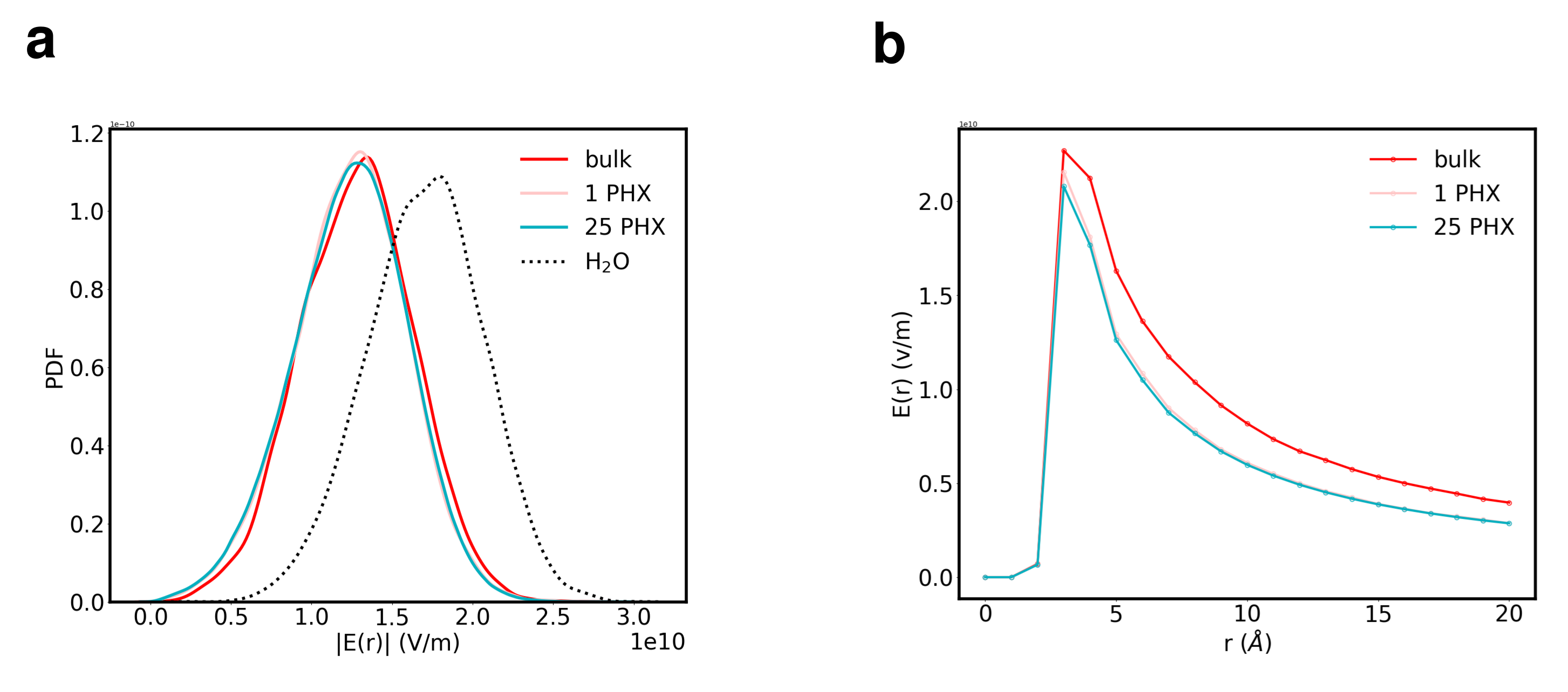}
  \caption{a) Probability density distribution of the magnitude of the electric field at the midpoint of the O-H bond of randomly selected water molecules (dotted curve) and of phenol molecules in all the three systems studied in this work (solid lines).
  b) Magnitude of the electric field generated by the water molecules at the midpoint of the O-H bond of the phenol molecules for the three systems studied as a function of the distance to this bond. As expected the decay goes as $1/r^2$.}
  \label{fig:Edistrib}
\end{figure}

As indicated earlier, one of the main goals of this study is to investigate with classical empirical potentials, how the electric field on the phenol moiety is tuned and affected by environmental factors. To this end, we begin by first comparing how the electric field on an O-H bond in liquid water compares to that of the hydroxyl group of the phenol. Figure \ref{fig:Edistrib}a) illustrates the distribution of the electric field on the center of the O-H of a water molecule in the bulk. We observe that the average field is $(1.7 \pm 0.4) \times 10^{10}$ V/m consistent with previous studies examining electric fields in both bulk water using different water models \cite{Reischl20022009,skinner2004,skinner2008,skinner2018}, as well as in water around amino acids \cite{jong2018}.

Moving to the electric field on the phenol, we compared the behavior of the fields and the corresponding fluctuations along the hydroxyl group for a single phenol in the bulk, at the interface and a situation consisting of 25 phenol molecules at the surface which corresponds to a surface access of 0.7 molecules/nm$^2$ consistent with those probed in previous experiments by Tahara and co-workers. Curiously, we observe that the electric field distributions are essentially the same - the average field is $(1.2 \pm 0.4)\times 10^{10}$ V/m in all three cases. Besides, the electric field experienced by the phenol is 30\% smaller than the field experienced by a water molecule. Thus, in this particular case, the electric fields of the phenol at the air-water interface are not particularly different from those experienced by water molecules in the bulk and furthermore, they are also not modulated or affected by the creation of phenol clusters/aggregates at the surface of water.

In order to probe deeper into the origins of the electric field distributions previously discussed, we determined the electric field separated into its contributions arising from different water molecules as a function of distance from the phenol (see Figure \ref{fig:Edistrib} b). We observe that for all the three systems involving phenol in different environments, the electric field has a peak below 5\AA{} and subsequently decays and becomes weaker on a larger length scale. This indicates that it is essentially the water molecules within the first one or two solvation shells that contribute the most to the electric field on the phenol.


Figure \ref{fig:Edistrib} b indicates that the radially resolved electric field obtained along the O-H bond of phenol for water molecules in the bulk, is in fact, slightly higher compared to the two situations of the phenol. The molecular origin of this can be rationalized by examining the solvent coordination number around the O-H groups both in bulk water and phenol. In Figure S1 in the Supporting Information (SI), we illustrate the running coordination number around the oxygen atom of the hydroxyl group comparing the three scenarios. Here it is clear to observe that within the first 5\AA{}, water molecules in the bulk are very unsurprisingly, better hydrated compared to phenol. This is hardly surprising given that water molecules in the bulk can donate and accept two hydrogen bonds on average. On the other hand, in the case of phenol, it can only accept and donate one hydrogen bond. Thus, it appears as though the local solvation may be a key player in determining the electric field. Later, we will explore this coupling more quantitatively by using information theory techniques.

\begin{figure}[ht!]
  \centering
  \includegraphics[width=1.0\linewidth]{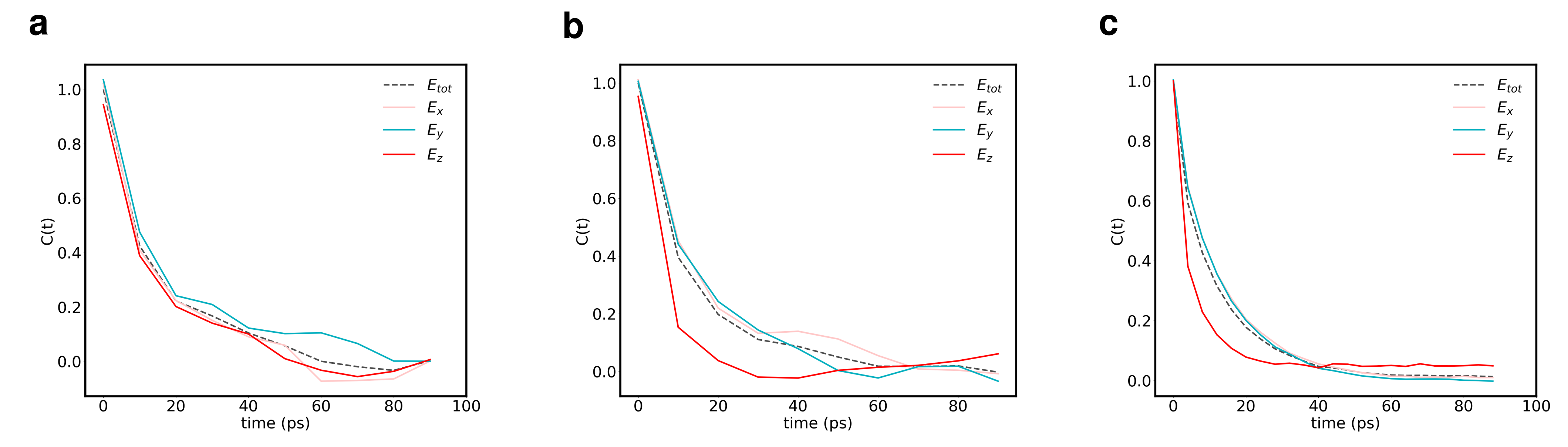}
  \caption{Autocorrelation function of the $\bar{E}$ decomposed in the three cartesian components and the total $|\bar{E}|$, for the three systems studied: a. bulk, b. 1PHX and c. 25PHX.}
  \label{fig:acf}
\end{figure}

The preceding discussion has focused on the static properties of the electric field and the magnitude of the contributions coming from different solvation layers. However, an equally important consideration in understanding the role of electric fields in chemical reactions is the timescale of field fluctuations. Specifically, if electric fields play a role in catalyzing chemical reactions, that implies that they would essentially be the slow collective variables. To this end, we examined the equilibrium time correlation functions C(t) for the electric field computed by equation \ref{eq:acf} and plotted in Figure \ref{fig:acf} resolving also the relaxation dynamics associated with the different components along the x, y and z directions.

\begin{equation}
\label{eq:acf}
       C(t) =  \left< \hat{E}(t)\cdot \hat{E}(0) \right>
\end{equation}

Figure \ref{fig:acf} shows the C(t) obtained for the three systems. Specifically, for phenol in the bulk, all the directions relax isotropically and can be fit with a single exponential corresponding to approximately 13 picoseconds. Moving to the single phenol at the air-water interface, we observe that the x and y directions relax with a single exponential with time constant of 10 ps. On the other hand, z component of the electric field projection decays faster (4 ps). Finally, for the case of the system with 25 phenols at the surface, we find that all the directions are fit by a biexponential which likely arises due to the heterogeneity induced by the presence of phenol clusters at the interfaces exposed to different amounts of water. As in the case of the one phenol molecule at the interface, the x and y directions relax on slower timescales (2 ps and 14 ps for the two components) compared to the z direction (1 ps and 7 ps for the two components).

All in all, the conclusions up to this point are that while there are certainly large electric fields experienced by the hydroxyl group of the phenol moiety at the air-water interface, these are not substantially different from the situation where the phenol is in the bulk or for that matter, the field on an O-H group of a water molecule in the bulk. These fields undergo large fluctuations - however, the timescales over which they relax occur on a timescale of 10 ps, which further shortens if one considers the field projection along the normal direction of the air-water interface typically referred as the interfacial field. Thus, the role of electric field fluctuations in promoting chemical reactions that would occur on much longer timescales requires further scrutiny. Next we move to examining specific correlations between the electric field and structural features that probe solvation.

\subsection{Solvation and Electric Fields}\label{subsec:solvation}

In the preceding sections, we have alluded to the fact that local solvation plays an important role in determining the electric fields experienced by the phenol moiety. In many ways, this is hardly surprising given that the electric field is determined by the charge density through Poisson's equation. Nonetheless, it is important to establish how these local fields and solvation are coupled. In this vein, we looked for correlations between the field and structural coordinates that are often used as a measure of solvation.

\begin{figure}[ht!]
  \centering
  \includegraphics[width=0.85\linewidth]{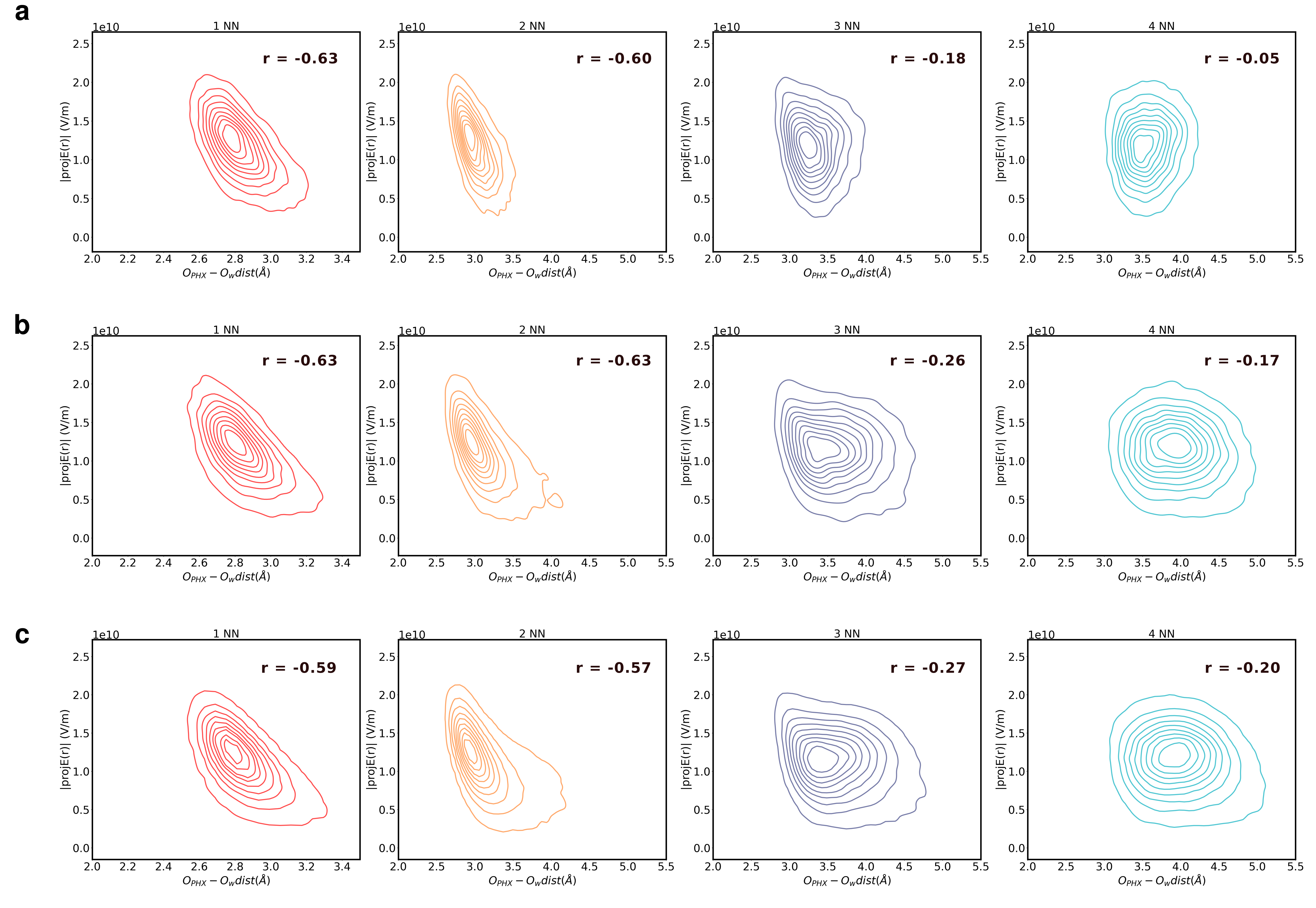}
  \caption{Density distributions showing the correlation between the $|proj\bar{E}|$ and the OO distance between the phenol ($O_{PHX}$) and the four nearest water molecules ($O_w$) for the a. bulk, b. 1PHX and c. 25PHX. The Pearson correlation coefficient (r) is showed in each panel.}
  \label{fig:Ecorr}
\end{figure}

More specifically, solvation of molecules in water is typically probed by examining the exposure or proximity of water molecules to the polar groups and the orientation of water molecules. To this end, we began by looking at the 20 nearest neighbor (NN)  water molecules to the hydroxyl group of the phenol for the three different systems. Figure \ref{fig:Ecorr} shows two-dimensional distributions between the magnitude of the electric field projected on the OH bond of the phenol ($|proj\bar{E}|$) and the four NNs. In all cases, we also computed the Pearson correlation coefficient which corresponds to the number in bold shown in each panel. Note that we used the projected electric field vector on the OH phenol since this group is mostly aligned with the field as shown in the SI in Figure S2 (the distribution of orientation for the OH bond of the phenol with respect to the z-axis is shown in figure S3).

Interestingly, for all the three systems the electric field displays a rather significant anti-correlation with the distance of the nearest neighbor water molecule. In other words, the stronger the hydrogen bond when the water-phenol distance gets shorter, the larger the field strength. This observation has been fully appreciated in previous works relating electric fields to vibrational spectroscopy - large fields along stronger hydrogen bonds result in larger red-shifts of the O-H stretch frequency\cite{eaves2005,cassone2019,Cassone2024}. Moving to the second NN, the correlation gets slightly weaker especially for phenol in the bulk (the correlation coefficient changes from -0.63 to -0.60). For phenol at the surface of water shown in the middle and bottommost panel, the correlation is also present although we begin to see some deviation from linearity. The coupling between the field and solvation appears to be significantly reduced for the third NN and is basically absent once we reach the fourth NN.

\begin{figure}[ht!]
  \centering
  \includegraphics[width=0.85\linewidth]{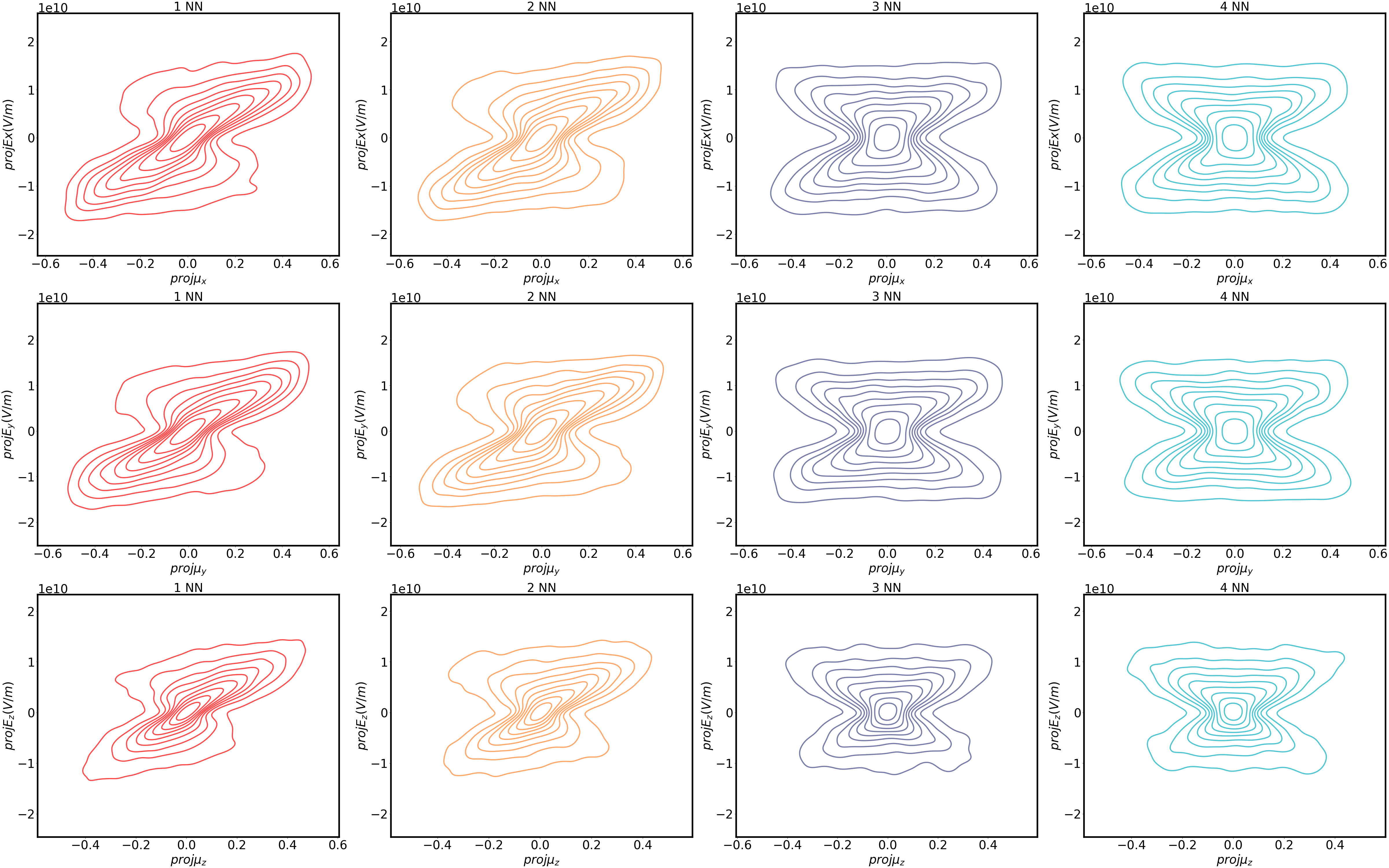}
  \caption{Density distributions showing the correlation between each cartesian component of the projected electric field ($projE_i$) and the corresponding component of the dipole vector of the four nearest water molecules ($\mu_i$) in the 1PHX system.}
  \label{fig:Edipcorr}
\end{figure}

Besides the radial proximity of a water molecule to the phenol, hydrogen bonding also involves specific directionality. For this reason, we examined the orientation of the water dipoles to the O-H vector of the phenol for the 4 NNs. In Figure \ref{fig:Edipcorr}, we illustrate the density plots showing the coupling between the field and the dipole vector of the 4 NN water molecules. Here specifically, we show two-dimensional density distributions involving the projection of both the field and dipole along the same direction (no cross-correlation was observed between variables, see Figure S3). Since the inferences are similar for the three cases, we focus on the single phenol at the interface. The analysis of the other systems is shown in the SI (Figures S3 and S4).

Figure \ref{fig:Edipcorr} shows that while the average field and dipole projections along all three directions are centered around zero, these two features exhibit a positive correlation along some directions of the density plot. This behavior is also observed for the 2NN but then disappears for the 3NN and 4NN. Thus, similar to the previous analysis examining the coupling between the field and the proximity of the close hydrogen-bonded waters, we observe that the two nearest neighbors appear to be playing the most critical role. Furthermore, the fact that the density plots associated with the 1NN and 2NN are skewed symmetrically with respect to the projection of the electric field along the O-H covalent bond of the phenol moiety implies that, in a flexible (or \emph{ab initio}) model, such a bond would experience field fluctuations which produce both stretching and contraction phenomena of similar magnitude. Due to the breaking of symmetry in the z-direction due to the presence of the air-water interface, the correlation plots for the z-components shown in the bottommost panel also display the same trends but are not exactly the same as those in the X and Y direction. In particular, the observed reduced spread of the density distributions shown in the bottommost panel of Figure \ref{fig:Edipcorr} mirrors the confinement effect along the z-axis involving reduced solvation. This aspect once again magnifies the strong correlation between the concepts of solvation and electric fields.

To conclude the observations of this section, it is clear through the visual representation of 2 dimensional density plots, that there is, as expected, a coupling between the electric fields experienced by the hydroxyl group on the phenol and solvation. One of the limitations of the preceding analysis is that it involves a dimensionality reduction examining the coupling only between pairs of variables. However, both the electric field and solvation coordinates (proximity and orientation) in principle encode several degrees of freedom which cannot be examined independently. We tackle this challenge in the next section.


\subsection{Information Imbalance: Linking Electric Fields and Solvation}\label{subsec:ii}

We have shown in the previous section, that local solvation collective variables such as the proximity of the water to the hydroxyl group of the phenol as well as the orientation are strongly coupled to the electric field. The relationships between the electric field and solvation were determined through visual inspection and by computing correlation coefficients. However, it is clear that the fluctuations of the phenol and its solvent environment involve the interaction of several degrees of freedom, for example solvation coordinates arising from both Figure \ref{fig:Ecorr} and \ref{fig:Edipcorr}.

Recently, our group has shown that fluctuations in aqueous systems typically occur in a high-dimensional manifold which need to be rationalized with information theoretical techniques\cite{Donkor_JCTC_2023,Donkor_JPCL_2024}. An example of this is a technique known as Information Imbalance\cite{Glielmo_PNASNexus_2022} (II) which at its core is a statistical test aimed at assessing the relative information content of different distance measures defined on the same data set. As we mention in the Methods section, the II has been extended to a differentiable formalism (DII) which allows for easily combining heterogeneous features and determine which are the most informative subsets of these features that describe a ``target'' data space. Referring to Eq.\ref{eq:dii}, for us to determine the most informative solvation-related collection variables that describe the electric field, we consider our ``target'' space, $d_B = \{E_x, E_y, E_z\}$, and our ``feature'' space, $d_A$ consists of a large set of solvation coordinates that we describe in more detail below. We define distances in the electric field space between points $i$ and $j$ (which are two different values of the electric field vector components at two times in our trajectory) as: $d_B^{ij} = ||E_x^i - E_x^j|| + ||E_y^i - E_y^j|| + ||E_z^i - E_z^j||$, where $||.||$ denotes the Euclidean norm. Distances in $d_A$ are also defined similarly using all the solvation coordinates.

\begin{figure}[ht!]
  \centering
  \includegraphics[width=0.40\linewidth]{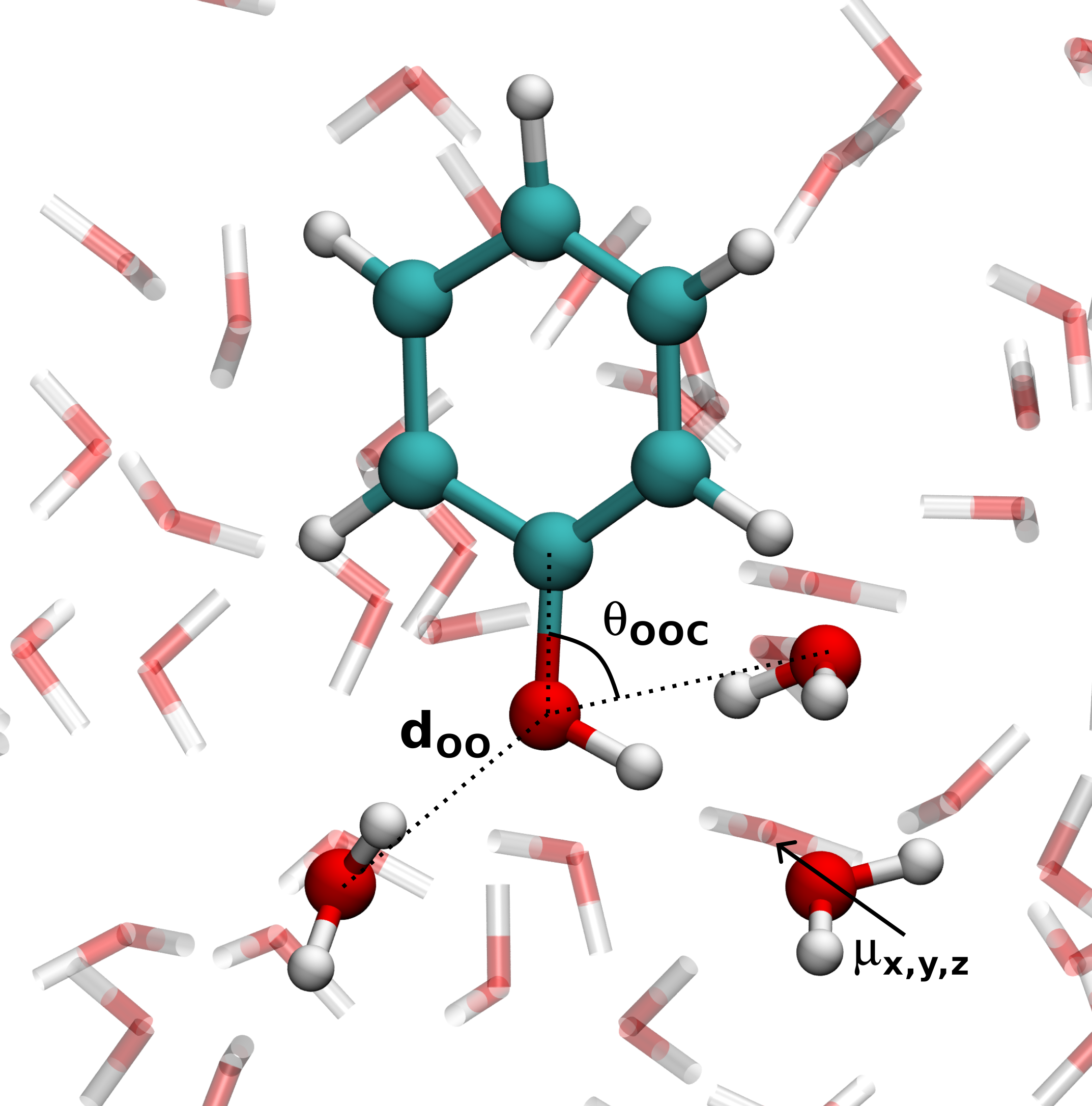}
  \caption{Structural variables used as feature set for the information imbalance algorithm. The same features were taken for the 20 nearest water molecules to the phenol (considered as the $O_{PHX}O_{water}$ distance).}
  \label{fig:features}
\end{figure}

We focused on a particular set of features which are visually illustrated in Figure \ref{fig:features}. For the 20 nearest water molecules to the hydroxyl group of the phenol, we extract several physical variables probing solvation. These include the distance, orientation of these waters relative to the O-H group (characterized by the dipole vector) and finally, the angle formed by these waters and the C-O bond of the phenol ($\theta_{OOC}$). This yields a total of 100 features with which we seek to find the combination that best predicts the electric field vector along the O-H. The choice of 20 water molecules covers a length-scale of 12 \AA{} and therefore extends up to 2-3 solvation shells. Recall that the II takes on values between 0 and 1, the former implying a set of features that can perfectly predict the quantity of interest. Thus, in the context of our current work, we seek to identify which combinations of solvation features lead to the lowest II.

\begin{figure}[ht!]
  \centering
  \includegraphics[width=0.85\linewidth]{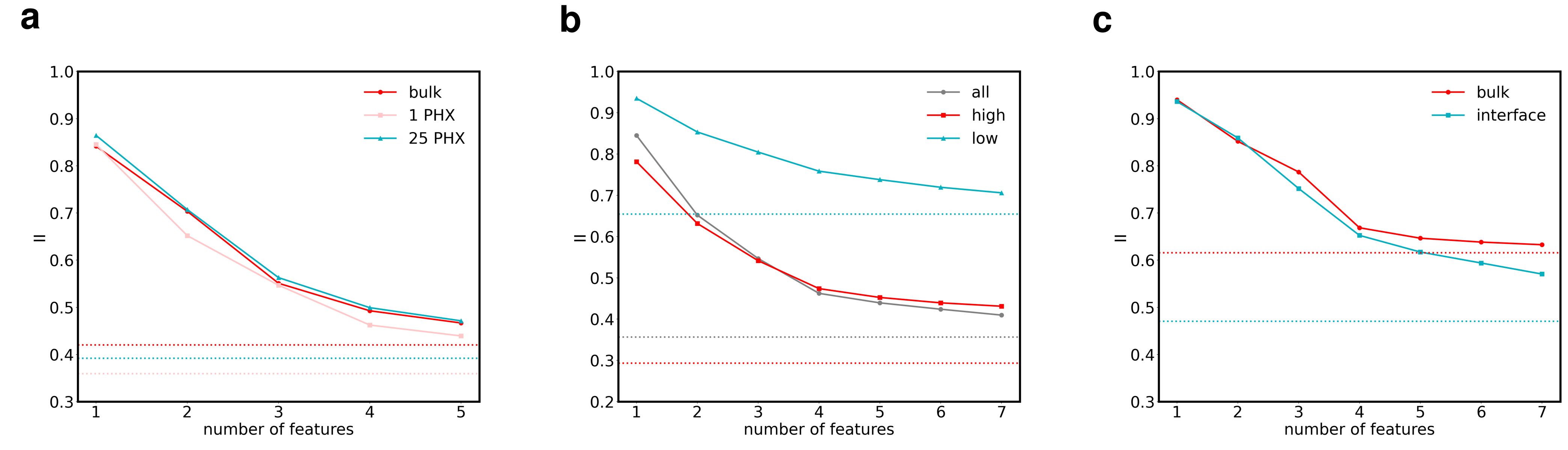}
  \caption{a. Information imbalance between the $|proj\bar{E}|$ and a set of structural features for phenol molecules in the 3 analyzed system. b. The same information imbalance but for a substet of high and low fluctuations of the $|proj\bar{E}|$ for the 1PHX system. c. II analysis for water molecules in the bulk and at the interface.}
  \label{fig:IIB}
\end{figure}

Using all the 100 features for the three systems, we obtain a II of approximately 0.4 (horizontal dotted lines in Figure \ref{fig:IIB}). Interestingly, we find that we need only 5 solvent features to reach this minimum II of 0.4 as seen in all the three curves of Figure \ref{fig:IIB} a. Table \ref{tab:table3} summarizes the best combination of features selected by the DII going from the best single to quintuplet combination for all the three systems. Interestingly, the best features for the phenol in the bulk appear to be those associated with the orientation of the water dipoles relative to the O-H of the phenol. For the two other cases involving the phenol at the interface, it appears as though the best features selected are variables associated with the proximity of 1NN or 2NN. We note, however, that in the case of predicting the electric field for the phenol in bulk, the II decreases quite marginally when going from the best 4 to the best 5 features (by about 0.02, see SI Table S1), and these 4 features are all related to the aforementioned orientation of the water dipoles.
Thus, it is primarily the same solvent features that play a role in determining the electric field both for phenol in the bulk and at the air-water interface.

Previous works\cite{Donkor_JCTC_2023,Donkor_JPCL_2024,Wild_SciRep_2024,Wild_NatComm_2025,deltattoRobustApproachInfer2025} have shown that getting values below 0.7 is usually considered a robust sign of having variables that are very informative on the target space. While it is possible to get an ``ideal'' II value close to 0.0 in model systems, which is the limiting case when our target space is fully described by our selected features, in real-world data sets, this is very unlikely. The presence of noise in the data, along with hidden variables that may better describe the target space but were not considered, leads to a deviation from the condition of perfectly informative features. That being said, however, as we see in Figure \ref{fig:IIB}, the II reduces from a value close to 1.0 (which represents the limiting case with uninformative features), to 0.4 as we add more meaningful variables. This allows us to determine the solvation coordinates that best predict the electric field vector.


\begin{table*}
\caption{\label{tab:table3}Features sequentially selected by the II algorithm in each system.}
\begin{ruledtabular}
\begin{tabular}{cccccc}
System & 1-Best feature & 2-Best features & 3-Best features & 4-Best features & 5-Best features \\
 \hline
bulk & [$\mu_{1,x}$] & [$\mu_{1,x}$,$\mu_{1,y}$] & [$\mu_{1,x}$, $\mu_{1,y}$, $\mu_{1,z}$] & [$\mu_{1,x}$, $\mu_{1,y}$, $\mu_{1,z}$, $\mu_{2,z}$] & [$\mu_{1,x}$, $\mu_{1,y}$, $\mu_{1,z}$, $\mu_{2,z}$,$\theta_{OOC,1}$] \\
1PHX & [$d_{OO,2}$] & [$d_{OO,2}$,$\mu_{1,x}$] & [$d_{OO,2,}$, $\mu_{1,x}$,$\mu_{1,y}$]  & [$d_{OO,2}$, $\mu_{1,x}$,$\mu_{1,y}$,$\mu_{1,z}$]   & [$d_{OO,2,}$, $\mu_{1,x}$,$\mu_{1,y}$,$\mu_{1,z}$,$\mu_{2,x}$]        \\
25PHX & [$d_{OO,1}$] & [$d_{OO,1}$,$\mu_{1,x}$] & [$d_{OO,1}$, $\mu_{1,x}$,$\mu_{1,y}$]  & [$d_{OO,1}$, $\mu_{1,x}$,$\mu_{1,y}$,$\mu_{1,z}$]   & [$d_{OO,1}$, $\mu_{1,x}$,$\mu_{1,y}$,$\mu_{1,z}$,$\mu_{2,x}$] \\

\end{tabular}
\end{ruledtabular}
\end{table*}

One of the arguments that has been floated in the literature regarding the role of electric fields on chemical reactivity at air-water interfaces is that it is their \emph{fluctuations} that need to be taken into account and that this could be a key driver in catalysis\cite{gordon2025,Hao2022}. To assess the role of these fluctuations, we repeat the DII analysis focusing on the data-space where the $|proj\bar{E}|$ arises exclusively from the tails of the distribution. We focused specifically on the regime that was greater than approximately $\pm 1.5 \sigma$ of the distribution, in other words, electric fields less than $6\times 10^9$ and greater than $1.6\times 10^{10}$ V/m. From this constrained dataset we repeated our analysis looking for the best set of features that leads to an optimal prediction of the electric field which is shown in Figure \ref{fig:IIB} b) (the values of the II corresponding to this plot can be found in Table S2). Interestingly, we observe that using the high-field region of the data (fields greater than $1.6\times 10^{10}$ V/m), the II plateaus at a slightly lower value when using the entire 100 features compared to what is observed in the left panel of Figure \ref{fig:IIB}. We thus included two more features into the analysis, which extends the set of water molecules considered beyond the 4th NN up to the 11th NN (the complete set of best features can be found in Table S3). Looking at the correlation between the $|proj\bar{E}|$ and the distance to the 4 nearest water molecules in Fig. S8 (upper panel) in the SI, it becomes clear that values of the electric field from the tails of the distribution yields stronger correlations with the nearest neighbor water distances. This in turn leads to a lower value of the II. On the other hand, for the lower fluctuations part of the distribution (fields less than $6\times 10^9$) the II based on solvation coordinates, is much higher and the correlation of the $|proj\bar{E}|$ with the 4 nearest water molecules is less pronounced (see bottom panel of Fig. S8 in the SI). The intuition behind this is that when the phenol begins to detach from the surface, it is less solvent exposed and therefore will naturally experience weaker fields.

As a final comparison, we performed the same II analysis for the electric fields acting at the midpoint of the OH bond in water molecules located in the bulk and at the interface. To this end, we calculated the Gibbs dividing surface (GDS) in the 1PHX system and, for each frame, selected one random molecule within $GDS \pm 2$ \AA{} and one random molecule in the bulk region. For both cases, we evaluated the electric field at the midpoint of one OH bond. We then repeated the analysis carried out for phenol: projection of the field onto the OH bond, calculation of the correlation of the field with respect to OO distances (Fig. S10), and correlation with the dipole components (Fig. S11 and S12) of the four nearest neighbors using the II analysis shown in Fig. \ref{fig:IIB}c. For the II analysis we employed analogous descriptors as in the phenol case, with the only modification being the replacement of the angle definition from $\theta_{OOC}$ to $\theta_{OOH}$, where H corresponds to the hydrogen atom of the target molecule not used as the reference point for the field calculation (see Fig. S9). As shown in figure \ref{fig:IIB}, the II values are larger than those obtained for phenol. This difference can be attributed to the fact that the descriptors selected for the water analysis—chosen to mirror those used for phenol—do not fully capture the underlying physical phenomena in water. To validate this, we plotted the electric field magnitude against the distance to the four nearest oxygen atoms (analogous to Fig. \ref{fig:Ecorr}), as well as the electric field components against the corresponding dipole moment components of the four nearest water molecules (Figs. 9–12, SI). These results show that, in the bulk, the correlations with distance remain strong up to the four nearest neighbors, whereas at the interface they weaken already at the fourth neighbor, similarly to the case of phenol. Regarding correlations with respect to the dipole moment, in the bulk only the first neighbor shows a significant correlation, which then decays rapidly, while at the interface correlations persist up to the third neighbor before fading. Altogether, these findings indicate that the molecular origin of the electric fields acting on phenol and on water molecules is governed by different features in each case. This suggests that, beyond the macroscopic $1/r^2$ decay of the field with distance, specific solvation characteristics must be considered when analyzing the effect of electric fields on a given system.

\section{Discussion and Conclusions} \label{sec:discussion}

Moving water from bulk to microdroplet conditions has been shown to significantly accelerate a broad class of chemical reactions. The microscopic origins of this effect is thought to be induced by anomalous properties of the air-water interface. However, the exact origins of this enhanced reactivity remain hotly debated and whether they actually arise from intrinsic effects due to water at the surface, remains under serious scrutiny. Among the various leading arguments that have been put forward, the presence of large electric fields and particular solvation effects within the interfacial region have been suggested to be key players in the underlying source of accelerated reactions.

If one considers a chemical reaction involving the breaking of a covalent bond, the notion is that the electric field experienced along the putative reaction coordinate may act in a different manner both in terms of its magnitude and direction at the surface of water compared to the bulk. At the same time, for some reactions, partial solvation at the interface could stabilize or destabilize in different ways, the reactants, transition-states or products. Since the electric fields must originate to a large extent from the charge density induced by the solvent or more generally environment, the two factors are inextricably intertwined.

In this computational work, we take a step toward understanding the relationships between electric field fluctuations and solvation using phenol as a model system. Using classical molecular dynamics simulations of an empirical potential of phenol in the bulk, at the interface and under more concentrated conditions that probe cluster/aggregate environments, we show that the magnitude and fluctuations of the electric fields along the hydroxyl group of the phenol show very similar characteristics. Although our models present large field fluctuations up to $\sim2\times10^{10}$ V/m, the timescales associated with this are rather fast occurring within $\sim$10 ps. In the language of slow collective variables that determine activation barriers, these timescales seem rather short lived and may not be relevant for being the slow degree of freedom needed to activate the reaction.

To quantify the relationship between the electric field experienced by phenol and the solvation environment, we deploy an information theoretical test that measures how well the electric field can be predicted from solvation coordinates including proximity and orientation of water to the hydroxyl moiety of the phenol. Perhaps unsurprisingly, electric fields and solvation cannot be treated as independent physical effects. The electric field on the phenol is determined by the surrounding water molecules and this effect is simply a general feature found both in phenol in bulk solution as well as at the air-water interface. These role of solvation is even more pronounced for electric field fluctuations creating larger values from the mean.

While the current work has focused on a model system, phenol, to study the behavior and relationships between electric fields and solvation, we speculate that these observations are rather generic and apply to most chemical systems especially those that consist of at least one polar moiety. We thus posit that the notion of the importance of electric fields in enhancing chemical reactions has perhaps been over-sensationalized in the community, a circumstance possibly rooted on the evidence that strong \emph{externally applied electric fields} are capable of profoundly modifying reaction networks and free energy landscapes. In the case of hydrogen-peroxide for example, the Mishra group has shown experimentally that its apparent enhanced yields in microdroplets may not be related to it occurring at the air water interface \cite{mishra2022}. Although this matter remains controversial in the literature of the area. Whether this extends to other reactions, however, remains an open question. At the same time, it is also worth noting that surface potentials of interfaces extracted from classical empirical models are different compared to those treated with \emph{ab initio} methods\cite{kathamnn2008,kathmann2011,wang2025charge}. It would be interesting in the future to use the information theoretic techniques to examine the relationships between the electric fields and solvation arising from models explicitly including electronic degrees of freedom and polarization.

In summary, the current investigation not only highlights the inextricable connection between the notion of chemical environment and local electric fields, but also points out the deep difference existing between externally applied (static and homogeneous) electric fields -- which might produce catalytic effects -- and the spontaneous fields arising from organized matter which fluctuate in magnitude and orientation. The physical and chemical impact of intrinsic versus externally applied electric fields on chemical reactions are thus distinct and any comparisons should be treated with caution. It may also be the case, that the key player involved in enhancing catalysis under the action of large external fields are actually solvation coordinates. This would be an interesting topic to explore in future studies.



\section{Supplementary Material}
The Supplementary Material includes data on the best features selected by the information imbalance procedure and the II values. Additionally, it has supplementary figures showing the running coordination number of the O atom of the phenol and the O atoms of the water molecules and the angle between the O-H bond of the phenol and the electric field. It also includes supplementary figures showing density distributions between each cartesian component of the projected electric field and the corresponding component of the dipole vector of the four nearest water molecules for each system.

\begin{acknowledgments}
DB, MM, GDM, and AH thank the European Commission for funding on the ERC Grant HyBOP 101043272. SDP, DB, MM, GDM, and AH also acknowledge MareNostrum5 (project EHPC-EXT-2023E01-029) for computational resources.
\end{acknowledgments}


\section*{Data Availability Statement}
The data that support the findings of this study are available from the corresponding author upon reasonable request.


\DeclareSIUnit\angstrom{\text {Å}}
\pdfstringdefDisableCommands{\let\ce\relax}

\newcommand{\equalcontrib}{\textsuperscript{*}}



\clearpage
\beginsupplement

\section{Methods: validation of classical force fields}

In order to validate the description of classical force fields that do not include polarization and charge transfer effects we conducted QM/MM calculations of the single phenol molecule at the interface using the AMBER/ORCA combined scheme \cite{amberorca}. We run QM/MM dynamics at the DFT level of theory, using the CAM-B3LYP functional together with the 6-31G* basis set. To do this, we took 50 frames from the classical trajectory and run 1 ps of QM/MM dynamics, including only the phenol molecule in the QM region. We then run a single point calculation on the last frame of these QM/MM dynamics, obtaining the dipole vector and magnitude of the phenol from DFT the electronic density. We also calculated these quantities on the same structures using the classical charges taken from the GAFF force field. The magnitude of the dipole moment according to the DFT calculations is of $2.23 \pm 0.05$ D whereas using the classical charges we get a value of $1.85 \pm 0.02$ D, giving a 20\% underestimation of the classical model with respect to the DFT calculation. We also calculated the cosine distance of the dipole moment vector calculated with DFT and with the classical charges obtaning a value of $0.91 \pm 0.01$, this indicates that the direction of the dipole moment is well reproduced by the classical model. Based on these results we conclude that for the porpoise of studying the magnitude and the molecular origins of the electric field at at the middle point  of the OH bond of the phenol molecule the classical model describes well enough the system.

\section{Phenol Electric Field: Statics and Dynamics}

\begin{figure}[ht!]
  \centering
  \includegraphics[width=0.50\linewidth]{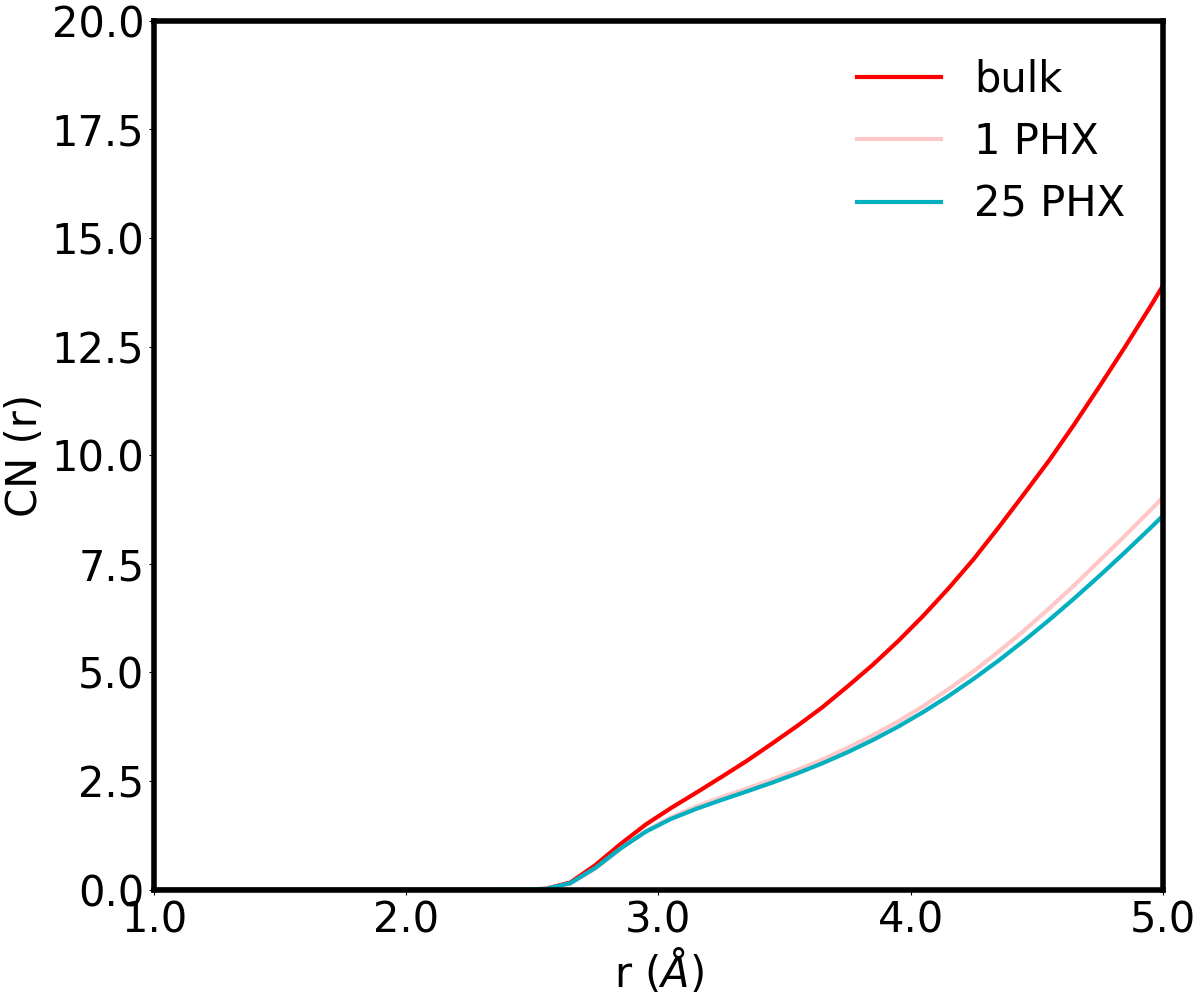}
  \caption{Running coordination number of the O atom of the phenol and the O atoms of the water molecules.}
\end{figure}

\begin{figure}[ht!]
  \centering
  \includegraphics[width=0.5\linewidth]{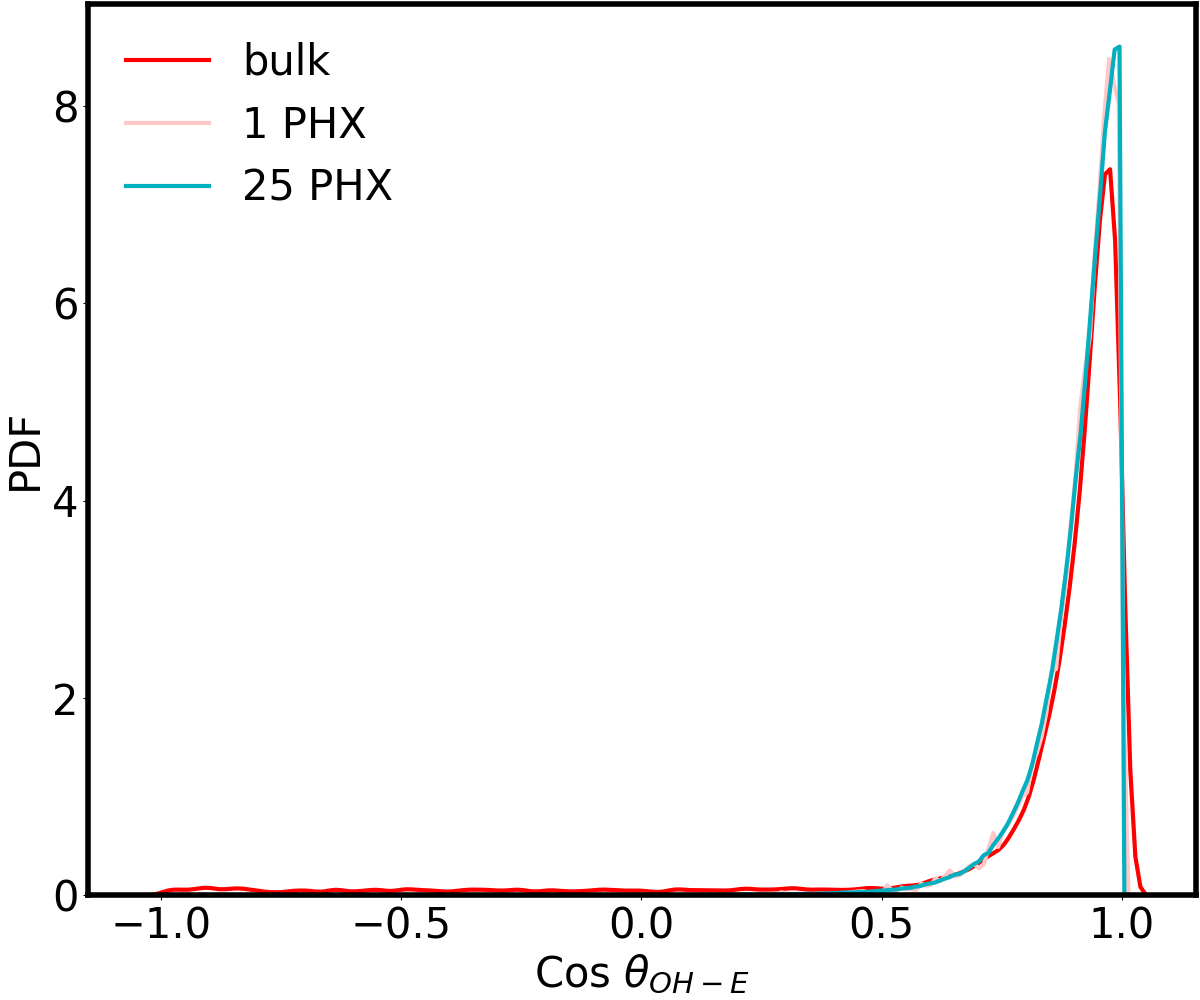}
  \caption{Angle between the OH bond of the phenol and the electric field. It can be seen that the OH bond is mostly aligned with the electric field.}
  \label{fig:angleOHE}
\end{figure}

\begin{figure}
    \centering
    \includegraphics[width=0.5\linewidth]{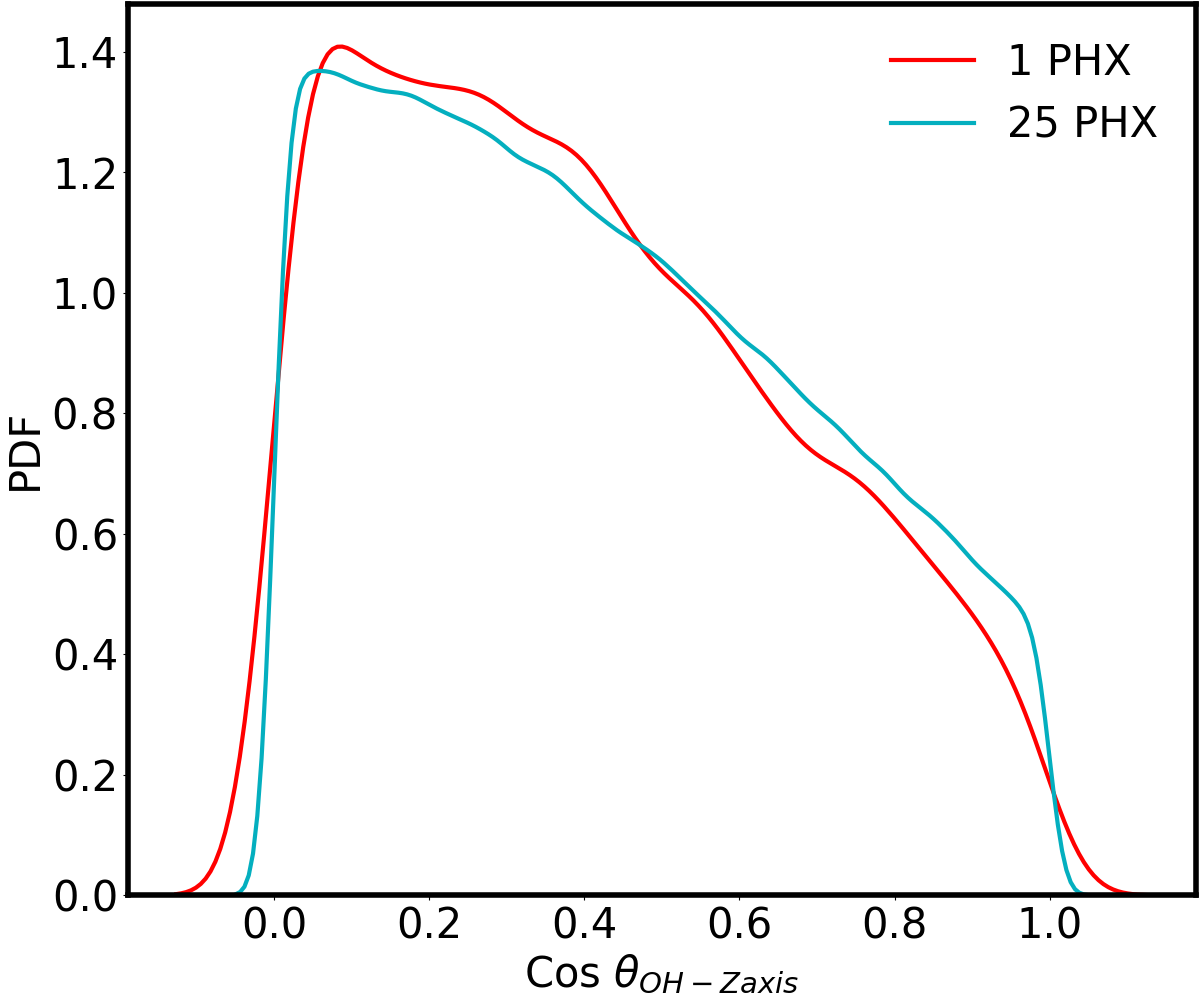}
    \caption{Distribution of orientations of the phenol OH bond with respect to the z-axis, the normal axis to the interface, for the 1PHX and 25PHX systems. The most probable orientation for the OH bond is perpendicular to the z-axis, being parallel to the interface.}
    \label{fig:ohdistrib}
\end{figure}

\clearpage

\section{Solvation and Electric Fields}

\begin{figure}[ht!]
  \centering
  \includegraphics[width=1.0\linewidth]{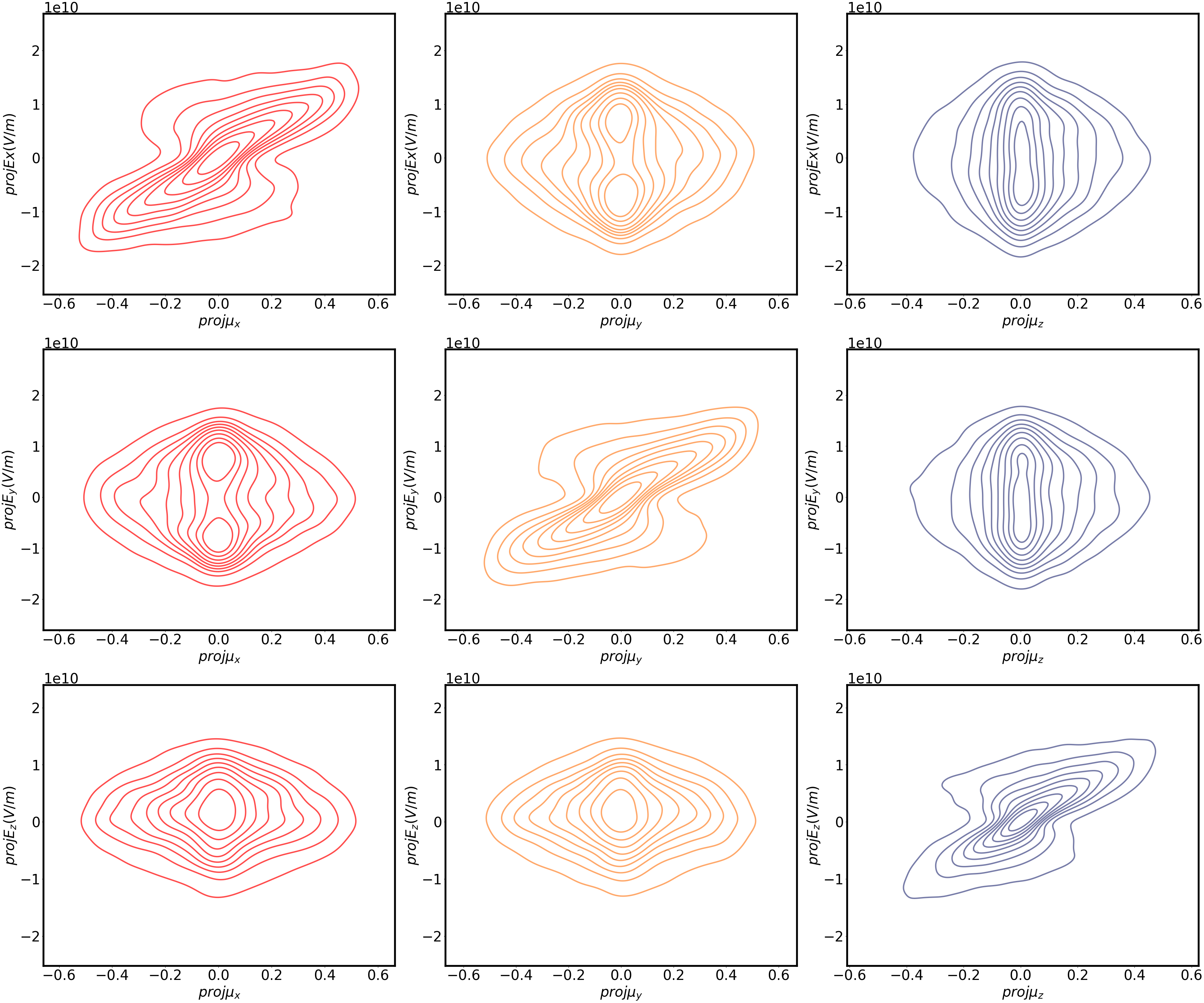}
  \caption{Density distributions showing the correlation between each cartesian component of the projected electric field ($projE_i$) and each component of the dipole vector of the first nearest water molecule ($\mu_i$ for 1NN) in the 1PHX system. Correlation is only observed for the corresponding components of $projE$ and $\mu$.}
  \label{fig:dipcorrbulk1}
\end{figure}

\begin{figure}[ht!]
  \centering
  \includegraphics[width=1.0\linewidth]{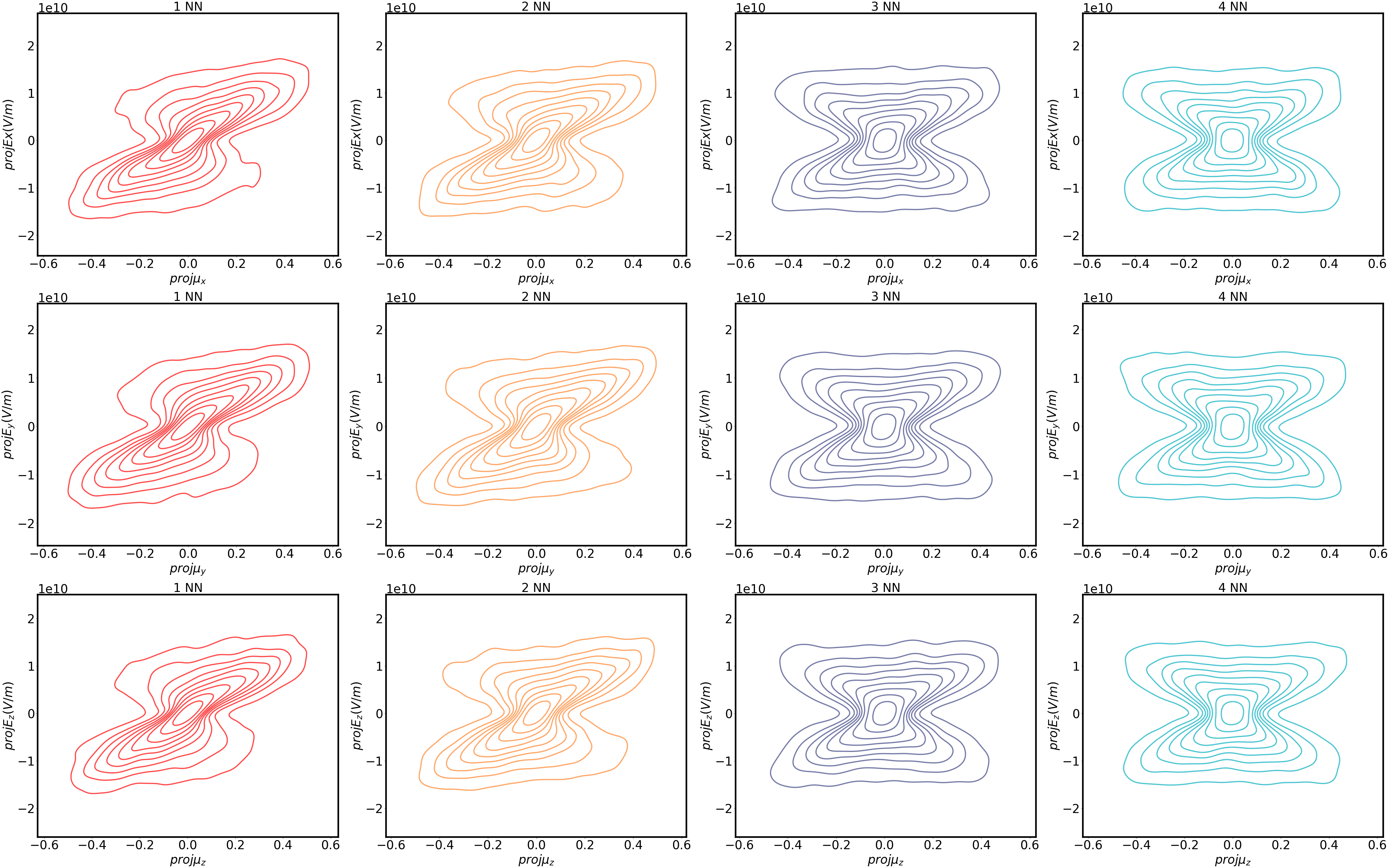}
  \caption{Density distributions showing the correlation between each cartesian component of the projected electric field ($projE_i$) and the corresponding component of the dipole vector of the first four nearest water molecules ($\mu_i$) in the bulk system.}
  \label{fig:dipcorrbulk2}
\end{figure}

\begin{figure}[ht!]
  \centering
  \includegraphics[width=1.0\linewidth]{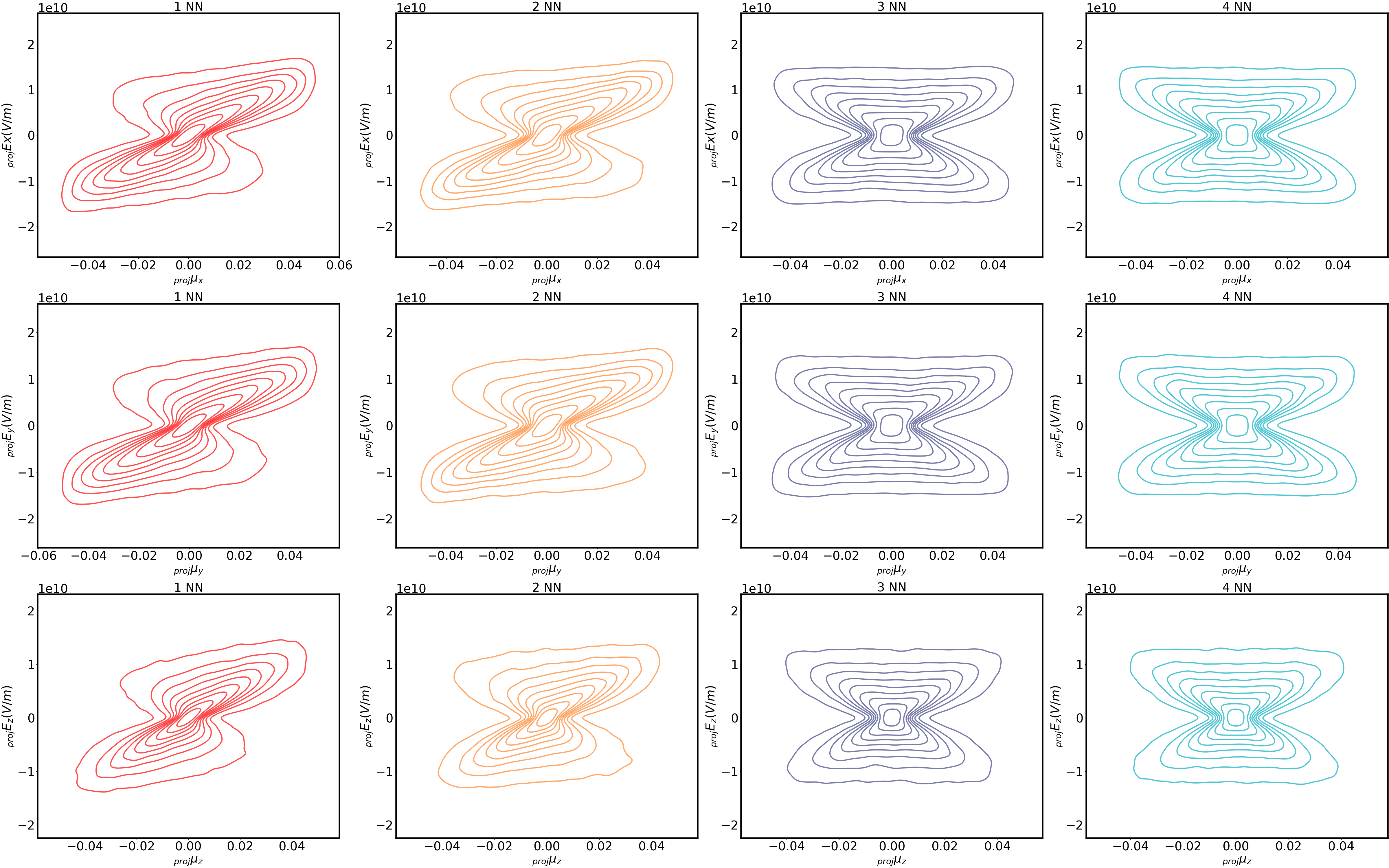}
  \caption{Density distributions showing the correlation between each cartesian component of the projected electric field ($projE_i$) and the corresponding component of the dipole vector of the first four nearest water molecules ($\mu_i$) in the 25PHX system.}
  \label{fig:dipcorr25phx}
\end{figure}

\begin{figure}[ht!]
  \centering
  \includegraphics[width=1.0\linewidth]{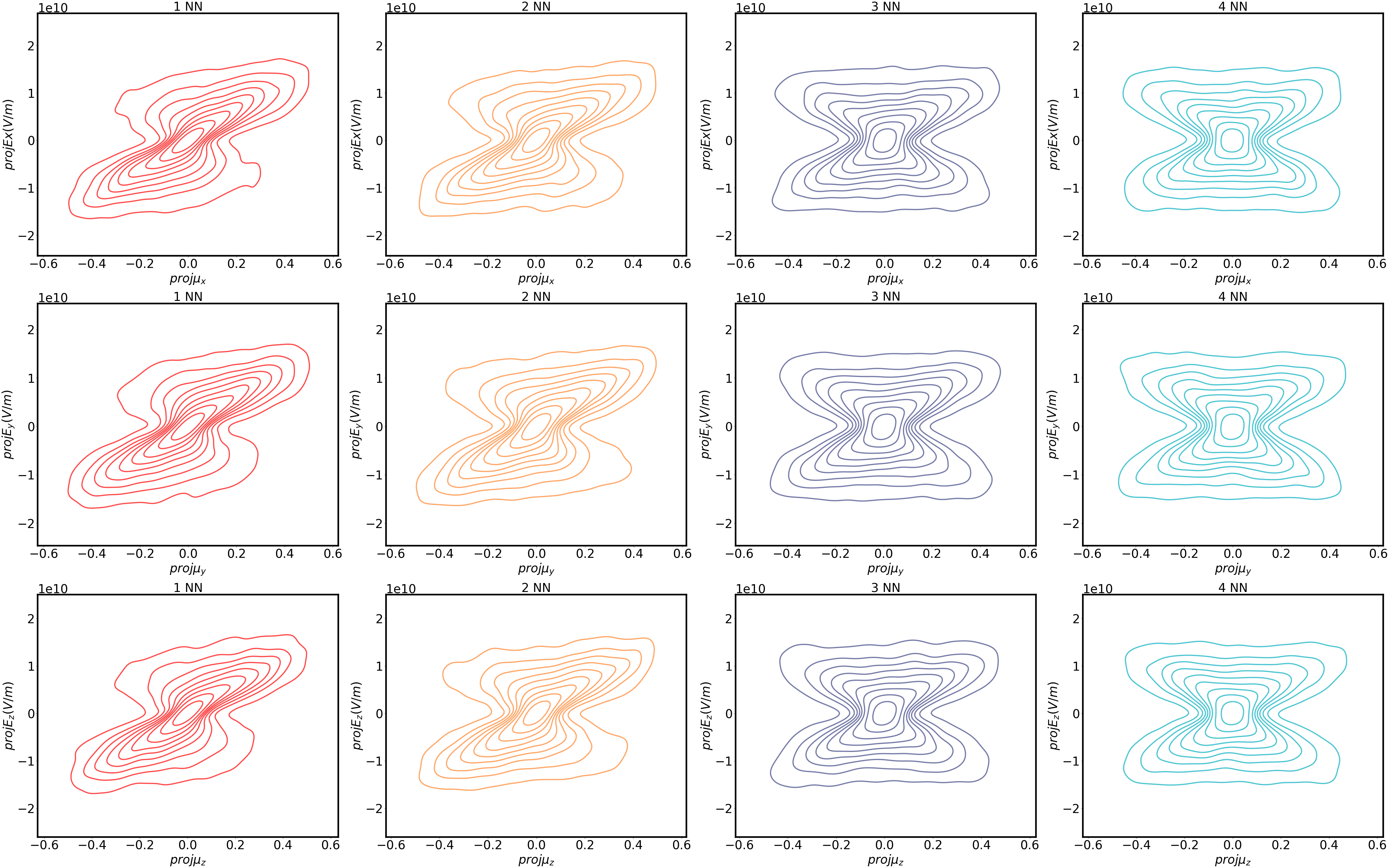}
  \caption{Density distributions showing the correlation between each cartesian component of the projected electric field ($projE_i$) and the corresponding component of the dipole vector of the first four nearest water molecules ($\mu_i$) in the bulk system.}
  \label{fig:corrfluct1}
\end{figure}

\begin{figure}[ht!]
  \centering
  \includegraphics[width=1.0\linewidth]{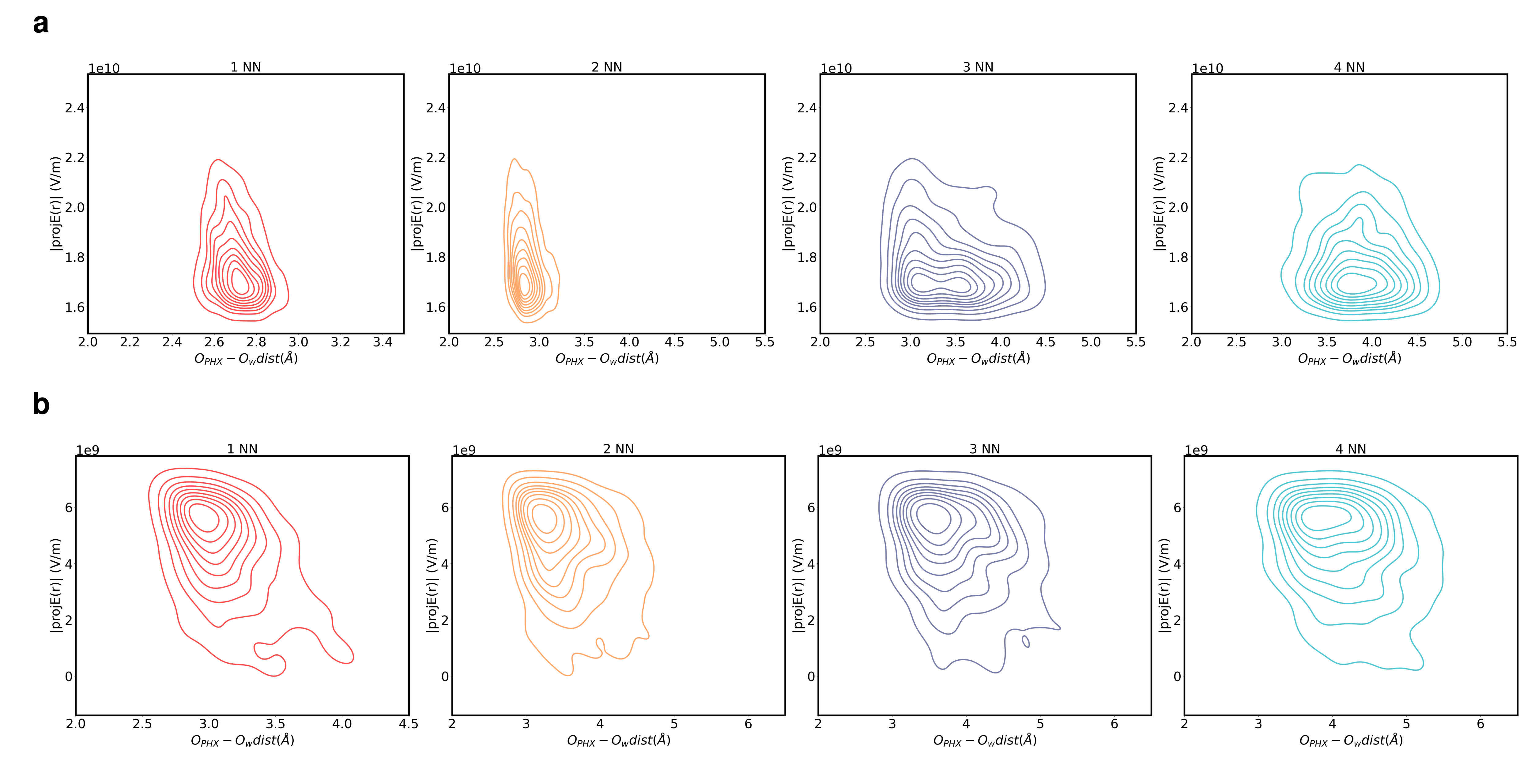}
  \caption{Correlation between $O_{PHX}O_w$ distance and the $|proj\bar{E}|$ for high (a) and low (b) fluctuations in the system with one phenol molecule at the interface (1PHX).}
  \label{fig:corrfluct2}
\end{figure}

\clearpage

\section{Information Imbalance: Linking Electric Fields and Solvation}

\begin{table*}[ht!]
\caption{II values according to the number of features selected for every system. The values correspond to the points in Fig.7.a.}
\label{tab:iivalues}
\renewcommand{\arraystretch}{1.3}
\begin{ruledtabular}
\begin{tabular*}{\textwidth}{c @{\extracolsep{\fill}} S[table-format=1.4] S[table-format=1.4] S[table-format=1.4]}

{Number of features} & {bulk} & {1PHX} & {25PHX}  \\
\midrule
1 & 0.8416 & 0.8452 & 0.8646 \\
2 & 0.7035 & 0.6520 & 0.7073 \\
3 & 0.5510 & 0.5468 & 0.5630 \\
4 & 0.4927 & 0.4624 & 0.4992 \\
5 & 0.4668 & 0.4394 & 0.4713 \\
\end{tabular*}
\end{ruledtabular}
\end{table*}

\begin{table*}[ht!]
\caption{\label{tab:iifluctuations}Features sequentially selected by the II algorithm for the subset of data corresponding to large fluctuations in the electric field for the 1 PHX system.}
\begin{ruledtabular}
\begin{tabular}{cccc}

System & high fluctuations & low fluctuations \\
\hline
1-Best features & [$\mu_{1,x}$] & [$\mu_{1,x}$] \\
2-Best features & [$\mu_{1,x}$,$\mu_{1,y}$] & [$\mu_{1,x}$,$\mu_{1,y}$] \\
3-Best features & [$\mu_{1,x}$,$\mu_{1,y}$,$\mu_{2,x}$] & [$\mu_{1,x}$,$\mu_{1,y}$,$\mu_{1,z}$] \\
4-Best features & [$\mu_{1,x}$,$\mu_{1,y}$,$\mu_{2,x}$,$\mu_{2,z}$] & [$\mu_{1,x}$,$\mu_{1,y}$,$\mu_{1,z}$,$\mu_{6,z}$, $d_{OO,3}$] \\
5-Best features & [$\mu_{1,x}$,$\mu_{1,y}$,$\mu_{2,x}$,$\mu_{2,z}$, $d_{OO,11}$] & [$\mu_{1,x}$,$\mu_{1,y}$,$\mu_{1,z}$,$\mu_{6,z}$, $d_{OO,3}$] \\
6-Best features & [$\mu_{1,x}$,$\mu_{1,y}$,$\mu_{2,x}$,$\mu_{2,z}$, $d_{OO,11}$,$d_{OO,4}$] & [$\mu_{1,x}$,$\mu_{1,y}$,$\mu_{1,z}$,$\mu_{6,z}$, $d_{OO,3}$,$\mu_{17,x}$]\\
7-Best features & [$\mu_{1,x}$,$\mu_{1,y}$,$\mu_{2,x}$,$\mu_{2,z}$, $d_{OO,11}$,$d_{OO,4}$, $\mu_{2,y}]$ & [$\mu_{1,x}$,$\mu_{1,y}$,$\mu_{1,z}$,$\mu_{6,z}$, $d_{OO,3}$,$\mu_{17,x}$, $\theta_{OOC,17}$]\\

\end{tabular}
\end{ruledtabular}
\end{table*}

\begin{table*}[ht!]
\caption{II values according to the number of features selected for every system. The values correspond to the points in Fig. 7.b in the main text.}
\label{tab:iivaluesfluct}
\renewcommand{\arraystretch}{1.3}
\begin{ruledtabular}
\begin{tabular*}{\textwidth}{c @{\extracolsep{\fill}} S[table-format=1.4] S[table-format=1.4] S[table-format=1.4]}
{Number of features} & {all} & {high} & {low}  \\
\midrule
1 & 0.8452 & 0.7815 & 0.9350 \\
2 & 0.6520 & 0.6312 & 0.8534 \\
3 & 0.5468 & 0.5416 & 0.8047 \\
4 & 0.4624 & 0.4739 & 0.7585 \\
5 & 0.4394 & 0.4524 & 0.7380 \\
6 & 0.4240 & 0.4394 & 0.7194 \\
7 & 0.4100 & 0.4312 & 0.7060 \\
\end{tabular*}
\end{ruledtabular}
\end{table*}

\begin{figure}[ht!]
  \centering
  \includegraphics[width=1.0\linewidth]{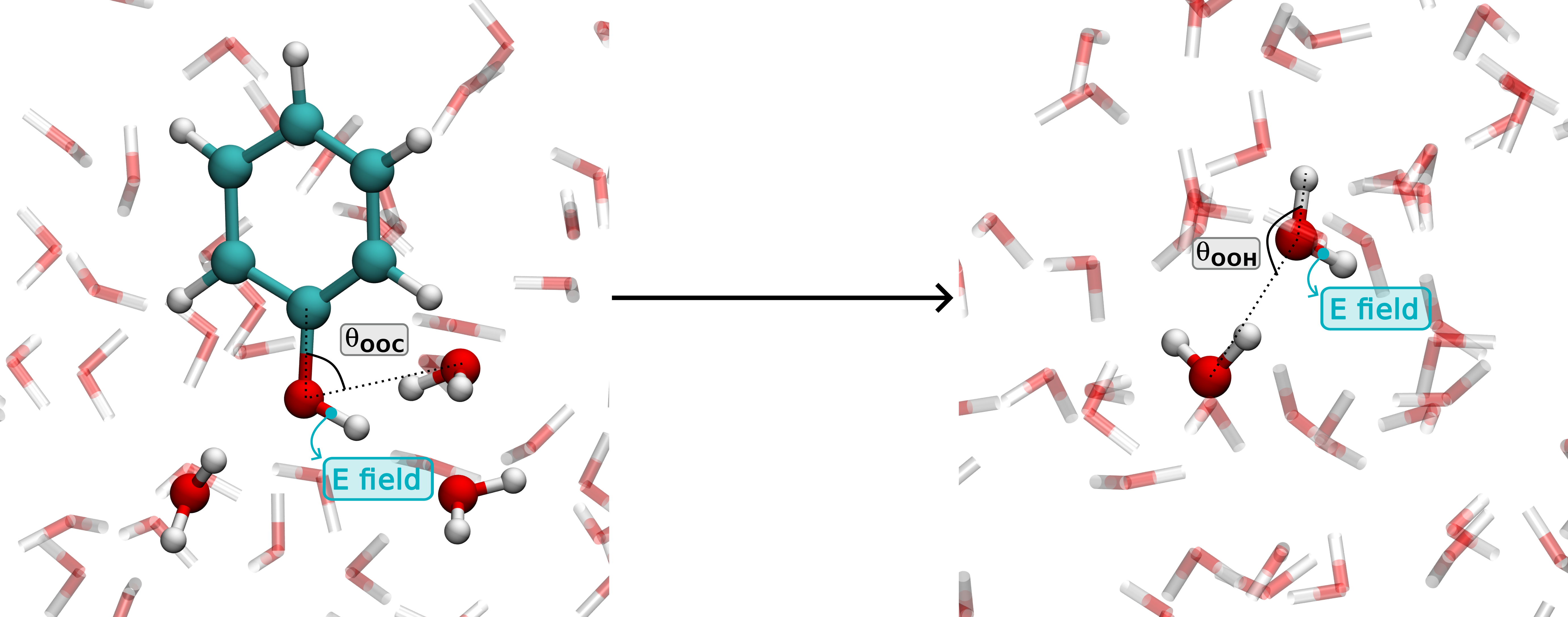}
  \caption{Angle used in the set of structural features used carry out the II analysis on water molecules. It was chosen to be analogous to the angle chose in the phenol case.}
  \label{fig:watfeat}
\end{figure}

\begin{figure}[ht!]
  \centering
  \includegraphics[width=1.0\linewidth]{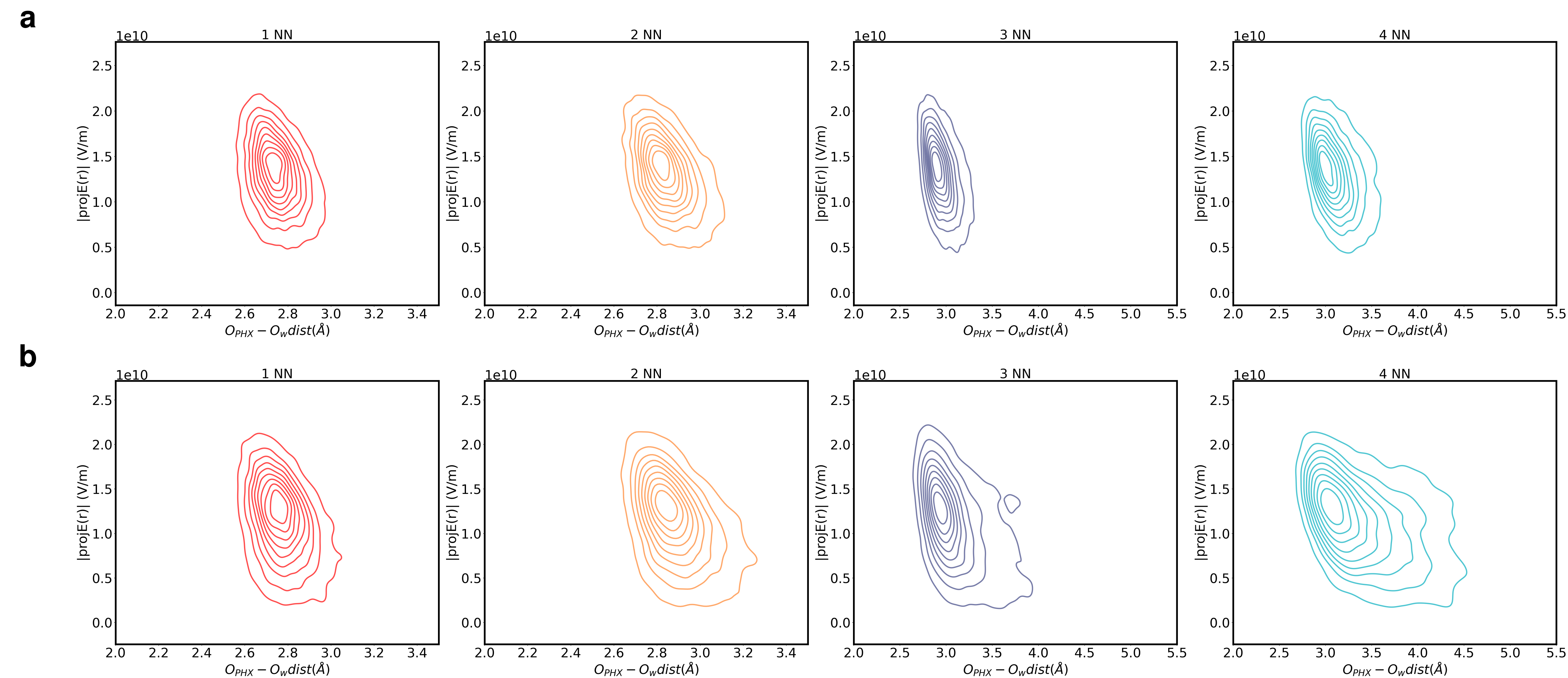}
  \caption{Correlation between $O_{PHX}O_w$ distance and the $|proj\bar{E}|$ for water molecules in the bulk (a) and at the interface (b).}
  \label{fig:corrfluct2}
\end{figure}

\begin{figure}[ht!]
  \centering
  \includegraphics[width=1.0\linewidth]{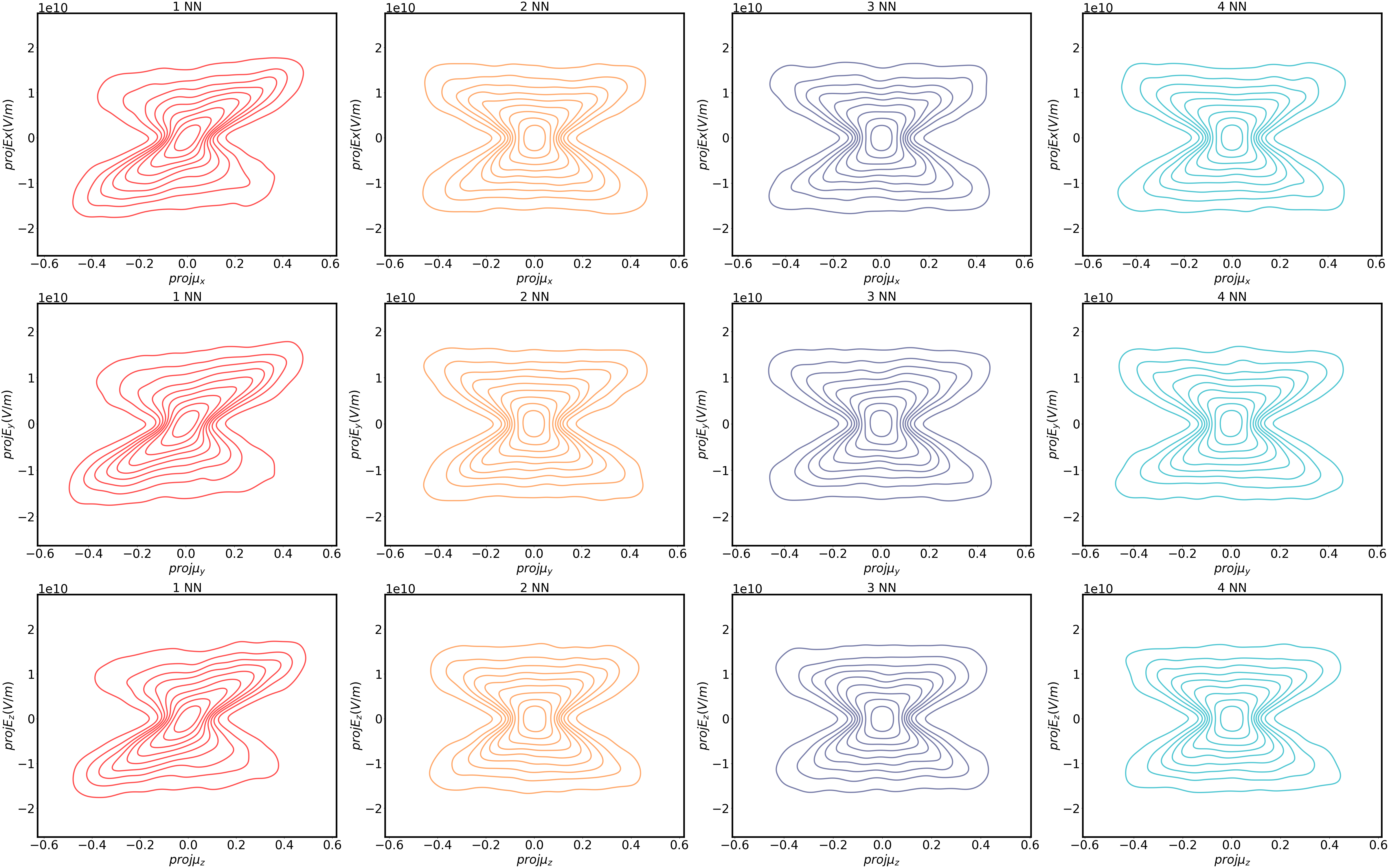}
  \caption{Density distributions showing the correlation between each cartesian component of the projected electric field ($projE_i$) and the corresponding component of the dipole vector of the first four nearest water molecules ($\mu_i$) for water molecules in the bulk.}
  \label{fig:corrfluct1}
\end{figure}

\begin{figure}[ht!]
  \centering
  \includegraphics[width=1.0\linewidth]{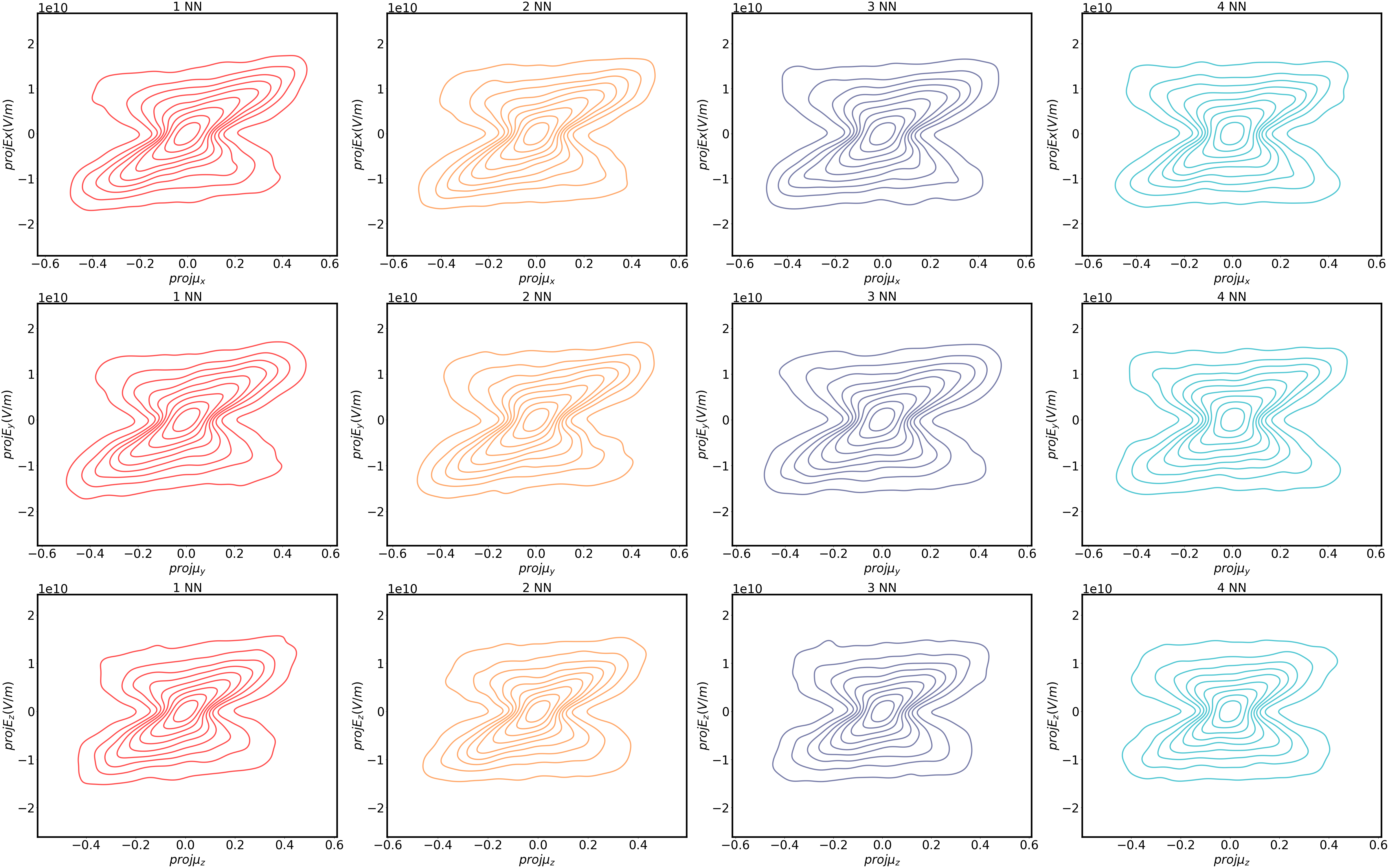}
  \caption{Density distributions showing the correlation between each cartesian component of the projected electric field ($projE_i$) and the corresponding component of the dipole vector of the first four nearest water molecules ($\mu_i$) for water molecules at the interface.}
  \label{fig:corrfluct1}
\end{figure}


\clearpage
\bibliography{refs}

\bibliography{refs}

\end{document}